\RequirePackage{lineno}
\documentclass[superscriptaddress,prd,twocolumn,showpacs,preprintnumbers,amsmath,amssymb,bibnotes,secnumarabic,altaffilletter,floatfix,aps,nofootinbib,10pt]{revtex4-1}

\usepackage{epsfig}
\usepackage{bigstrut}
\usepackage{amsmath}
\usepackage{graphicx}
\usepackage{epsfig}
\usepackage{fancyhdr}
\usepackage{calc}
\usepackage{lineno}
\usepackage{natbib}
\usepackage[usenames]{color}

\linenumberdisplaymath
\usepackage[colorlinks=true, breaklinks=true,pdfstartview=FitV, linkcolor=black, citecolor=black, urlcolor=black]{hyperref}

\newcommand{\btab}[1]{
  \begin{table}
  \begin{center}
  \begin{tabular}{#1} \hline
}

\newcommand{\btabn}[2]{
  \begin{table}
  \caption{#2}
  \begin{center}
  \begin{tabular}{#1}
  \hline\hline
}

\newcommand{\etabn}[1]{
  \hline
  \hline
  \end{tabular}
  \end{center}
  \label{t:#1}
  \end{table}
}

\newcommand{\btabns}[2]{
  \begin{table}
  \small
  \caption{#2}
  \begin{center}
  \begin{tabular}{#1}
  \hline\hline
}
\newcommand{\btabnss}[2]{
  \begin{table}
  \footnotesize
  \caption{#2}
  \begin{center}
  \begin{tabular}{#1}
  \hline\hline
}

\newcommand{\btabnsss}[2]{
  \begin{table}
  \scriptsize
  \caption{#2}
  \begin{center}
  \begin{tabular}{#1}
  \hline\hline
}

\newcommand{\bstab}[1]{
  \begin{table}
  \small
  \begin{center}
  \begin{tabular}{#1} \hline
}

\newcommand{\mtab}{
  \hline
  \end{tabular}
  \end{center}
}

\newcommand{\etab}[1]{
  \label{t:#1}
  \end{table}
}

\newcommand{\bfig}[2]{
  \begin{figure}
  \begin{center}
  \includegraphics[width=#2cm]{#1}
}

\newcommand{\blfig}[2]{
  \begin{figure}
  \begin{center}
  \includegraphics[width=#2cm,angle=270]{#1}
}

\newcommand{\efig}[1]{
  \label{f:#1}
  \end{center}
  \end{figure}
}

\newcommand{\mcc}[2]{\multicolumn{#1}{c}{#2}}
\newcommand{\mcl}[2]{\multicolumn{#1}{l}{#2}}

\newcommand{\bsb}{\bigstrut[b]}
\newcommand{\bst}{\bigstrut[t]}

\newcommand{\bstb}{\bigstrut[t]\bigstrut[b]}

\newcommand{\bCentre}{\begin{center}}
\newcommand{\eCentre}{\end{center}}
\newcommand{\be}{\begin{equation}}
\newcommand{\ee}{\end{equation}}
\newcommand{\bd}{\begin{displaymath}}
\newcommand{\ed}{\end{displaymath}}

\newcommand{\cref}[1]{Chapter~\ref{c:#1}}

\newcommand*\patchAmsMathEnvironmentForLineno[1]{%
      \expandafter\let\csname old#1\expandafter\endcsname\csname #1\endcsname
      \expandafter\let\csname oldend#1\expandafter\endcsname\csname end#1\endcsname
      \renewenvironment{#1}%
         {\linenomath\csname old#1\endcsname}%
         {\csname oldend#1\endcsname\endlinenomath}}%
    \newcommand*\patchBothAmsMathEnvironmentsForLineno[1]{%
      \patchAmsMathEnvironmentForLineno{#1}%
      \patchAmsMathEnvironmentForLineno{#1*}}%
    \AtBeginDocument{%
    \patchBothAmsMathEnvironmentsForLineno{equation}%
    \patchBothAmsMathEnvironmentsForLineno{align}%
    \patchBothAmsMathEnvironmentsForLineno{flalign}%
    \patchBothAmsMathEnvironmentsForLineno{alignat}%
    \patchBothAmsMathEnvironmentsForLineno{gather}%
    \patchBothAmsMathEnvironmentsForLineno{multline}%
}

\newcommand{\nnbarModels}{
\begin{figure}[h]
\includegraphics[width=0.5\textwidth]{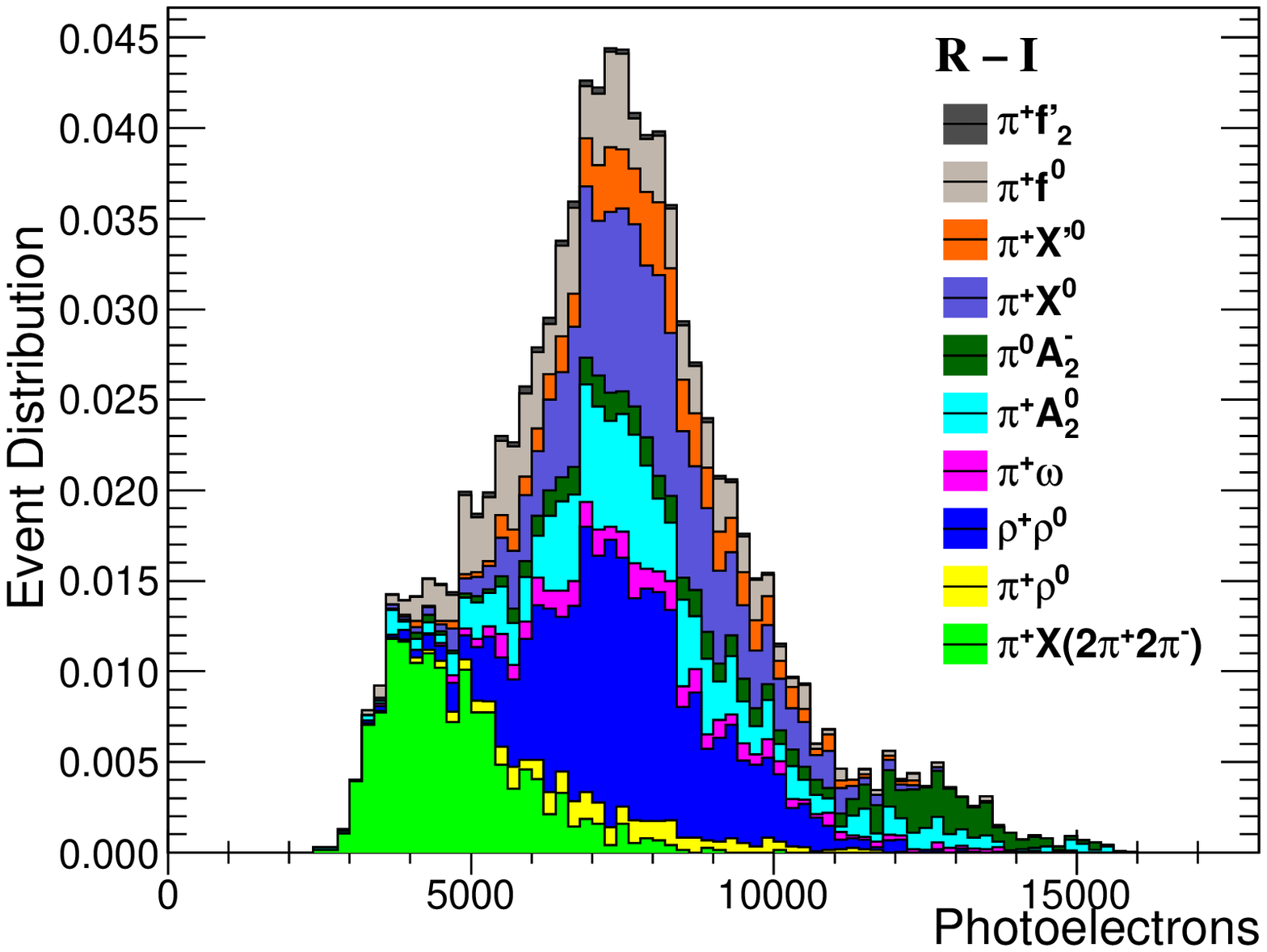}
\includegraphics[width=0.5\textwidth]{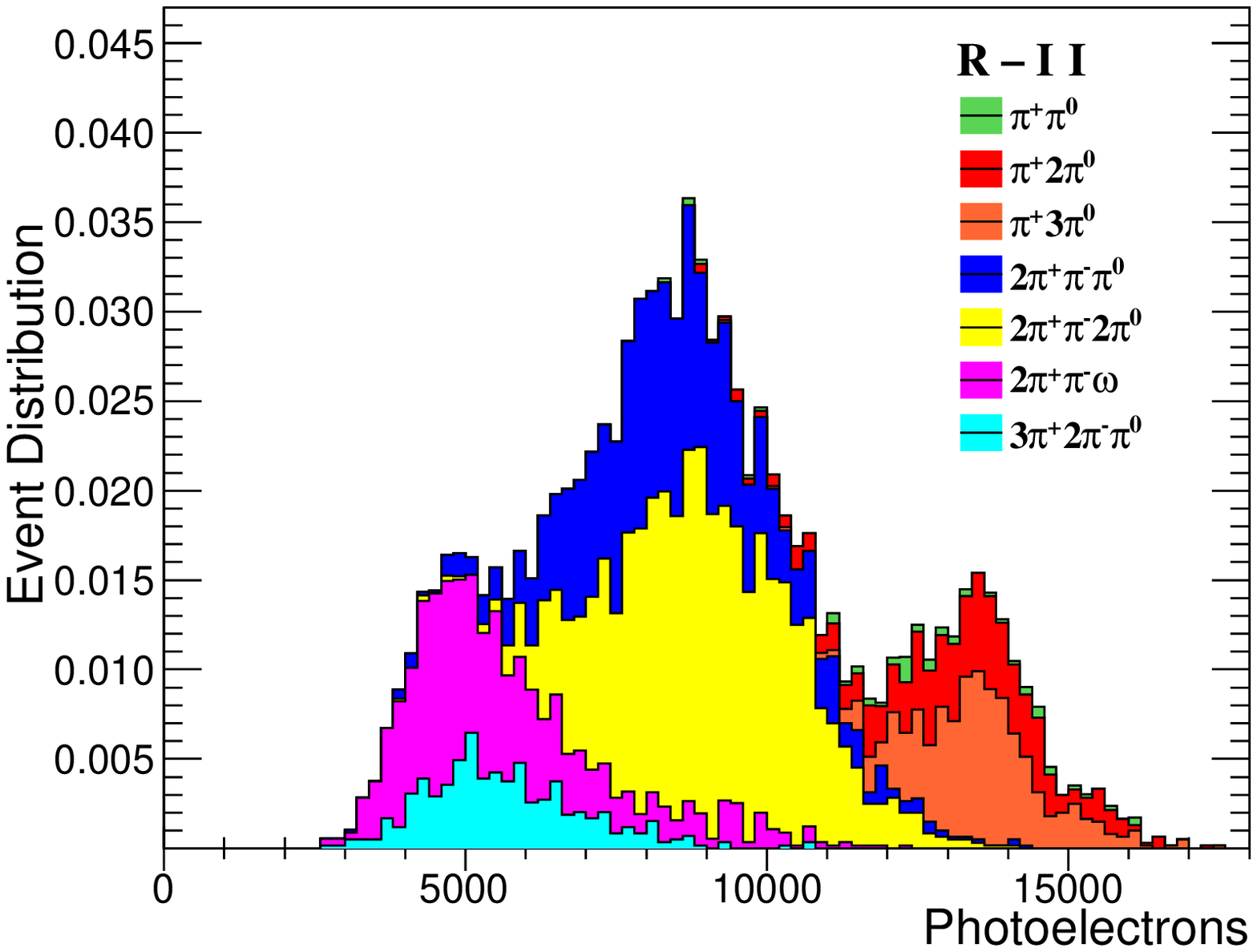}
\caption[]{Simulated SNO detector response in the two neutron-antineutron momentum regimes as described in Section \ref{SNO:Deuteron}.  Here the number of photoelectrons is proportional to the visible energy of the event. Using a conversion factor of 9 p.e./ MeV, $n-\bar{n}$ events have a visible energy signature in the range of 200~MeV to 1.9~GeV.}
\label{NNBAREnergy}
\end{figure}
}

\newcommand{\xnoedFigures}{
\begin{figure}[h]
\includegraphics[width=0.4\textwidth]{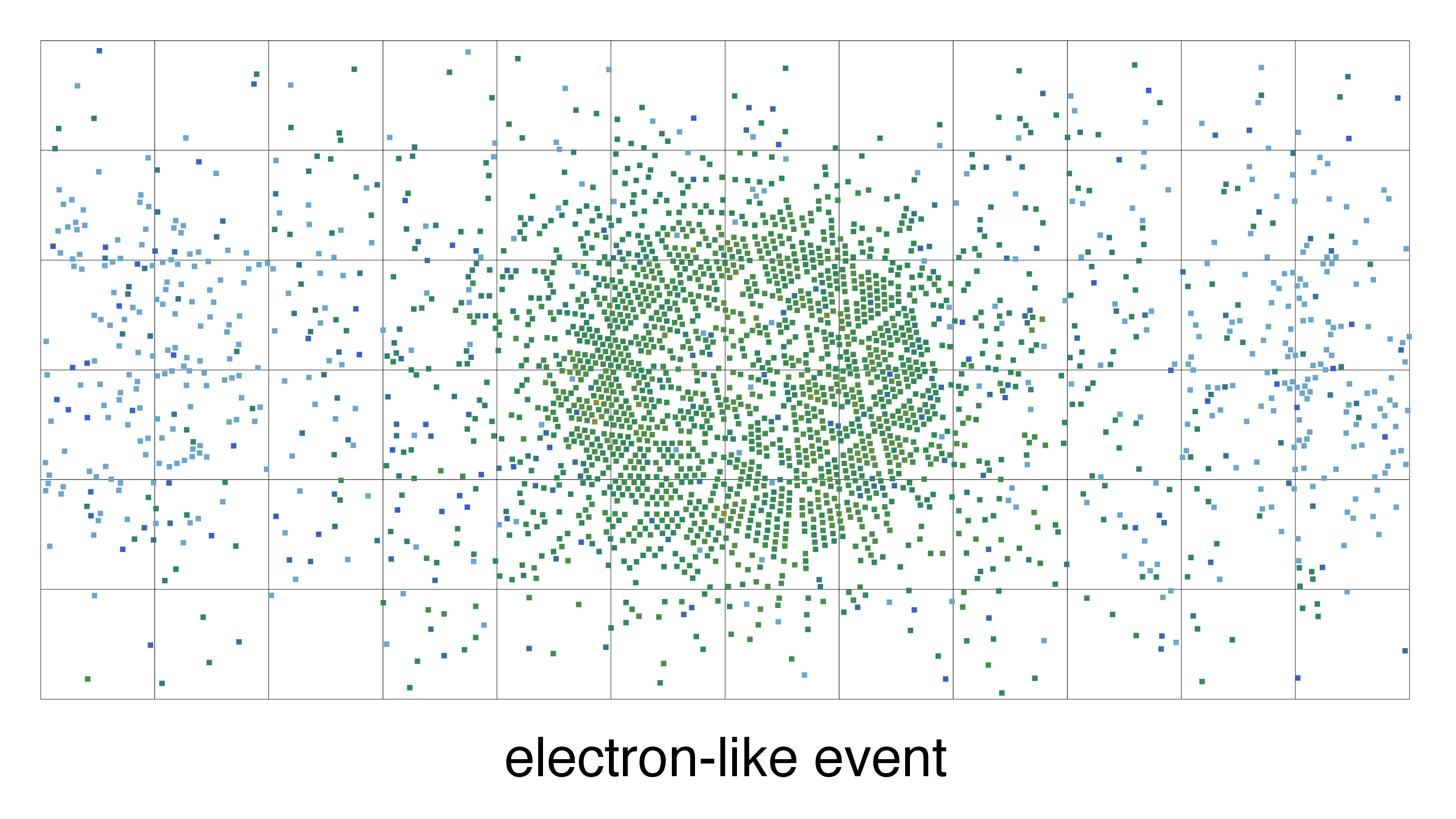}
\includegraphics[width=0.4\textwidth]{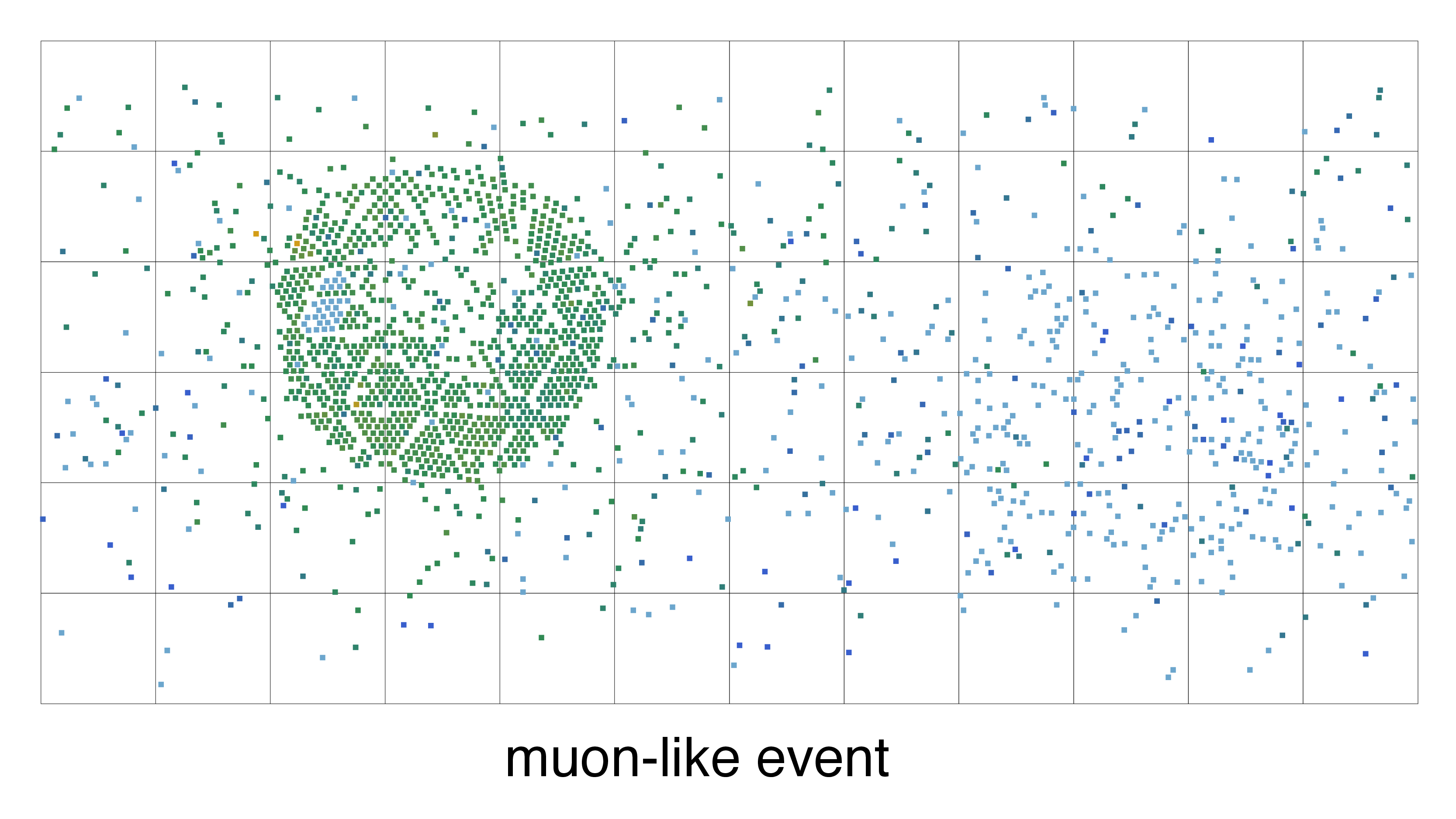}
\caption[]{Simulated Cherenkov electron ring (top) demonstrating the showering effect  and muon ring (bottom) demonstrating the absence of the showering effect. Both particles were generated with 600 MeV of kinetic energy and placed at the center of the detector; the resulting PMT hit pattern is projected in ($\cos\theta_{\rm PMT}$, $\phi_{\rm PMT}$) space.  The colors shown represent the time the PMT has been hit, where green points represent PMTs that were hit first and blue points are PMTs that were hit at a later time.
}
\label{CherenkovSignature}
\end{figure}
}

\newcommand{\dodecahedron}{
\begin{figure}[t]
\includegraphics[height=0.48\textwidth,angle=90]{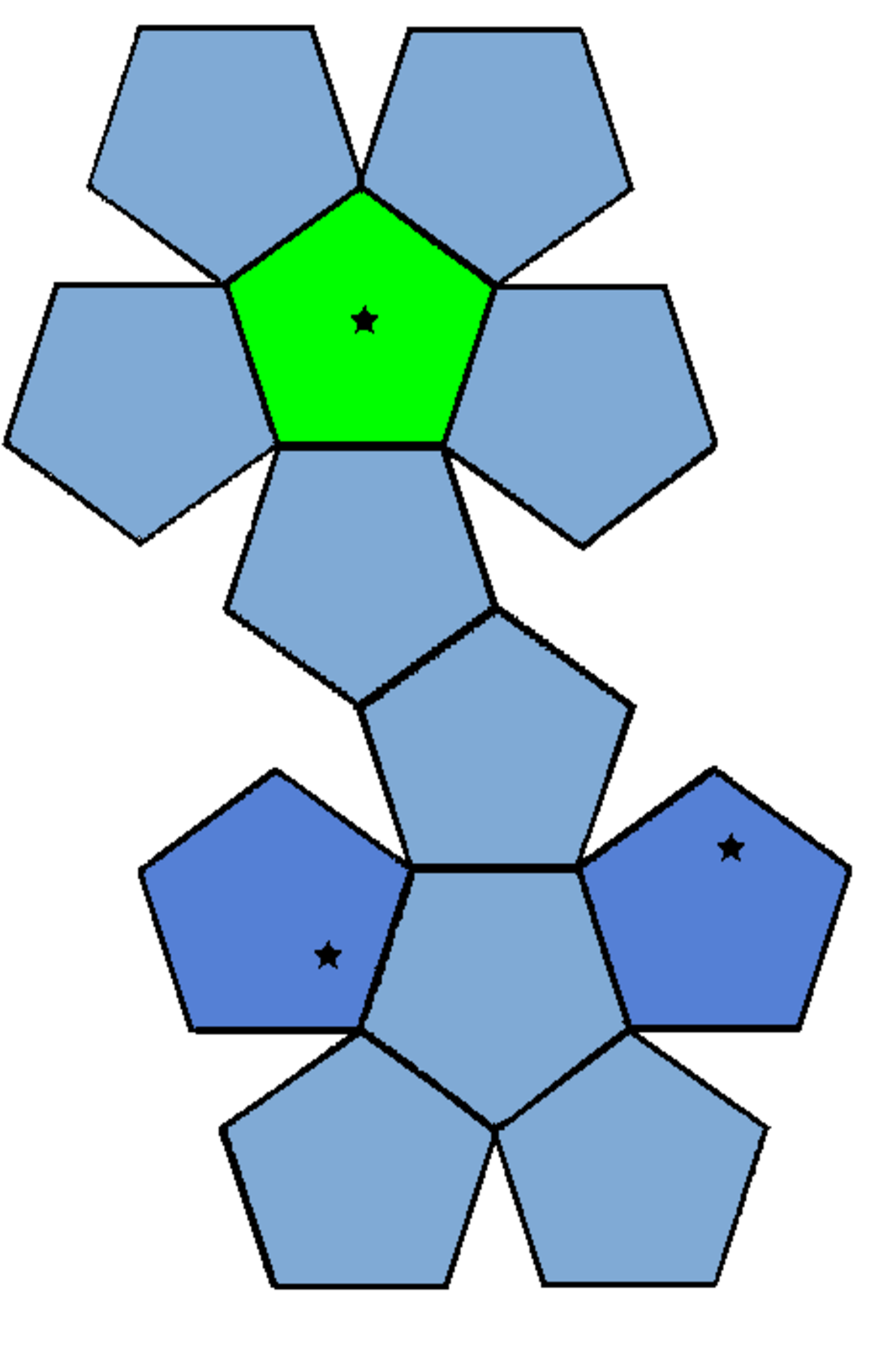}
\caption[]{An example of the dodecahedron construct implemented for multiple-ring detection in the ring-parameter space.  Each black star represents the ($\cos\theta_{mp}$, $\phi_{mp}$) coordinates of a local high density maxima of mid-points, which is denoted as ($\cos\theta^{\dagger}_{mp}$, $\phi^{\dagger}_{mp}$). The dodecahedrons structure is rotated and centered around the point in the ring-parameter space with the highest density, in this case the black star in the green region. In this example, this event is reconstructed as a three-ring event, the primary ring in the green section and two secondary rings in the dark blue sections.  The light blue sections are regions where no high density of mid-points is found.}
\label{Dodecahedron}
\end{figure}
}

\newcommand{\AngularDistribution}{
\begin{figure}[t]
\includegraphics[width=0.5\textwidth]{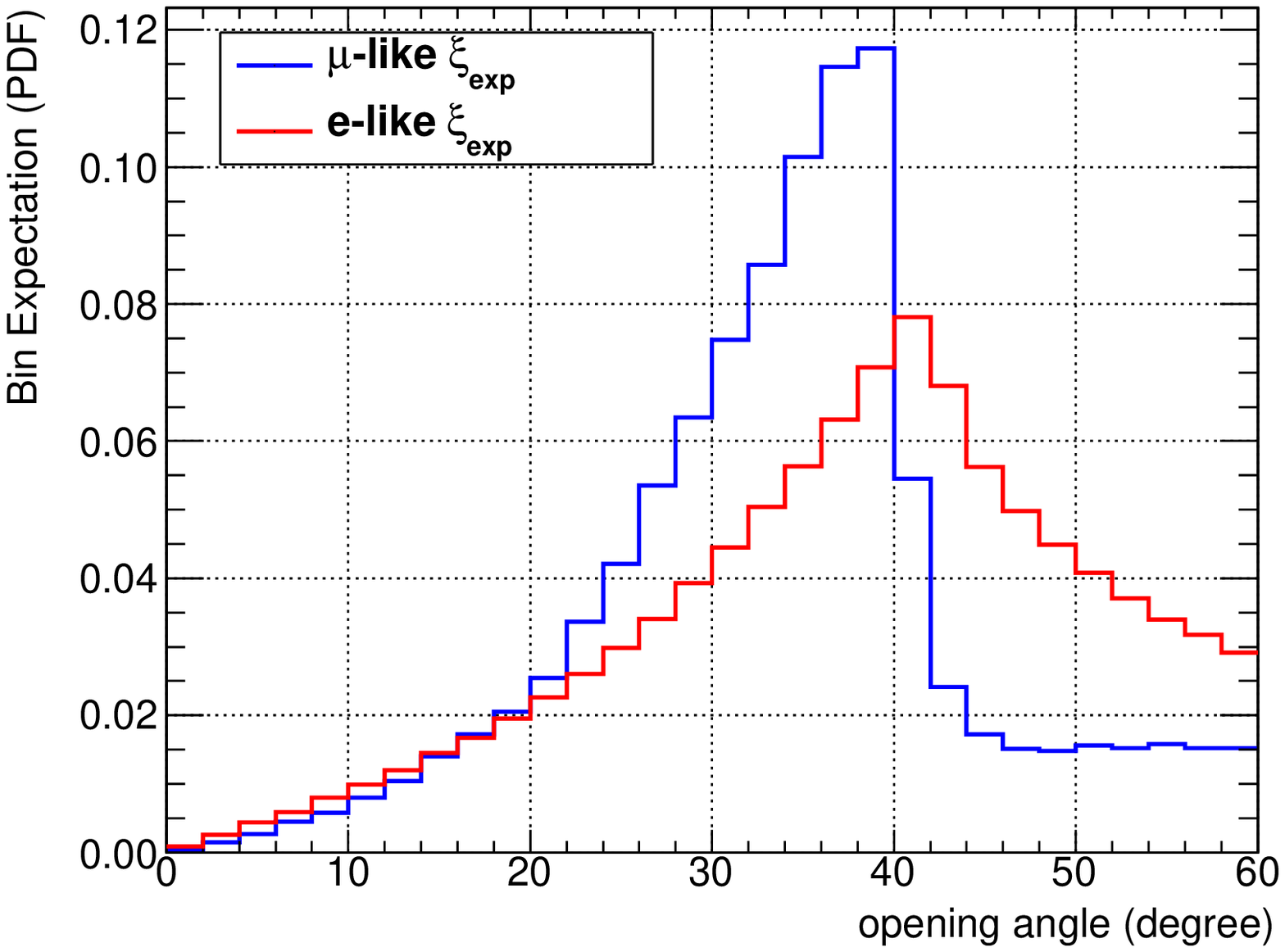}
\caption[]{The distribution of the opening angle subtended at the fitted vertex by the arc between the fired PMT and the ring center coordinate ($\cos\theta_{mp}^{\dagger}$, $\phi_{mp}^{\dagger}$) for both a showering ($e$-like) and non-showering ($\mu$-like) particle.}
\label{Angulardistribution}
\end{figure}
}

\newcommand{\piplusReconstructionFigures}{
\begin{figure}
\includegraphics[width=0.49\textwidth]{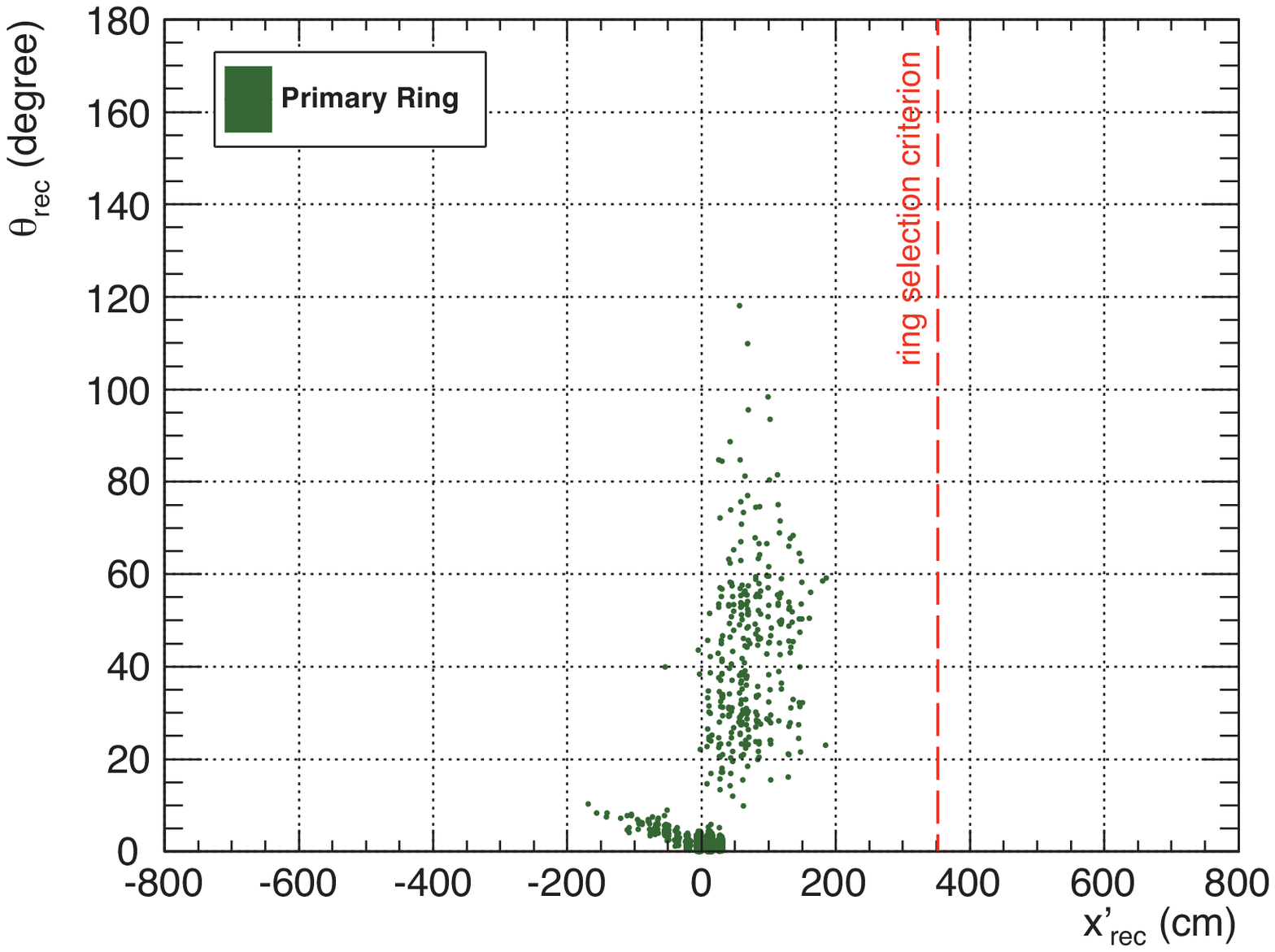}
\includegraphics[width=0.49\textwidth]{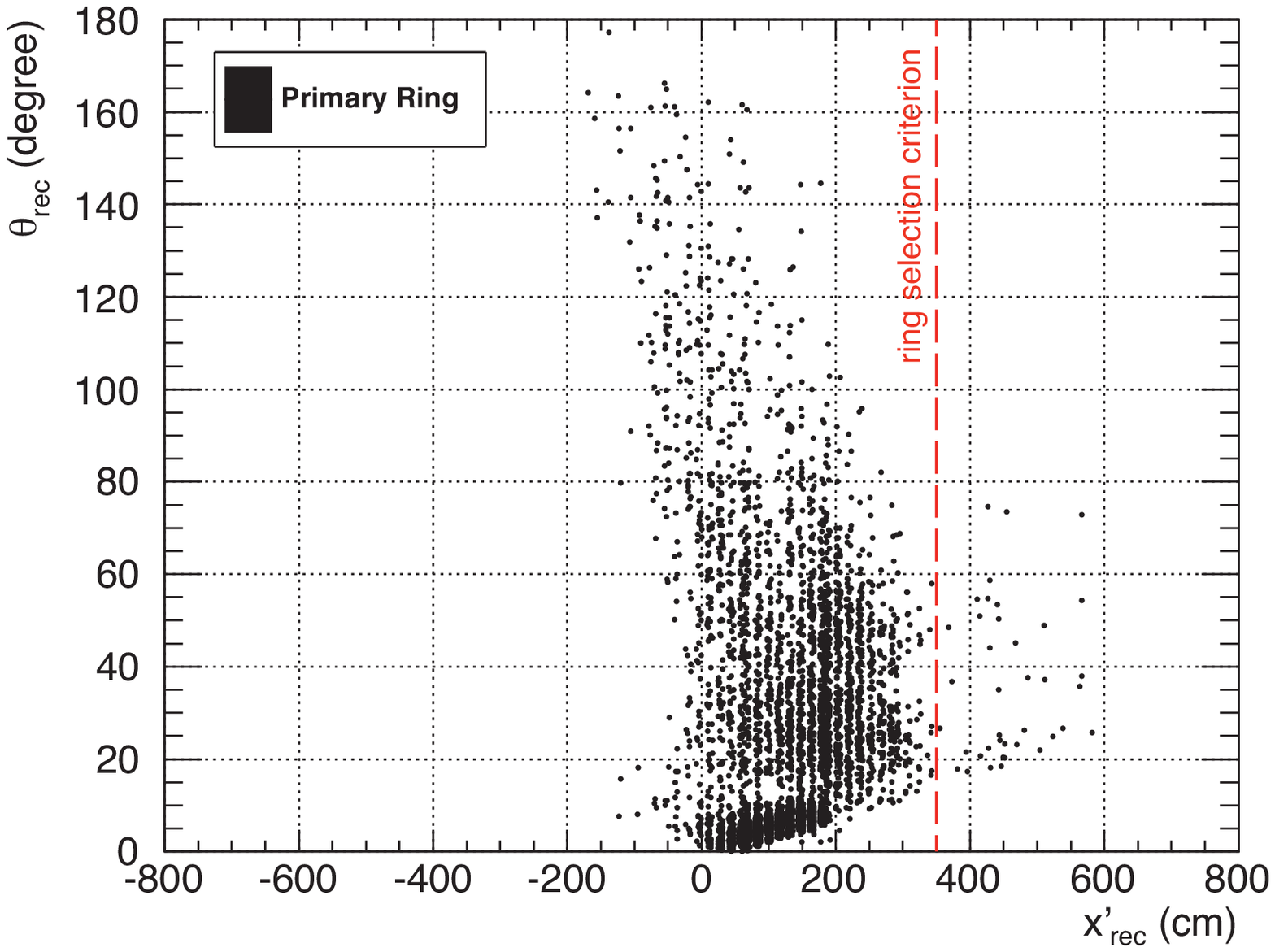}

\caption[]{Reconstruction of $\pi^+$ simulated at the origin of the detector using a non-showering (top) and showering (bottom) expectation, highlighting the complexity of single pion reconstruction. $x'_{rec}\hat{u}_{rec}$ is the reconstructed vertex position and $\theta_{rec}$ is the angle of the reconstructed track with respect to the original true track. Visual cross-verification of these events shows good ring recognition except for rings with  $x'_{rec}>350$~cm. This is used as a ring selection criterion. }
\label{fig:piplus_mu_reco}
\end{figure}

}

\newcommand{\xnoedPatho}{
\begin{figure}
\includegraphics[width=0.49\textwidth]{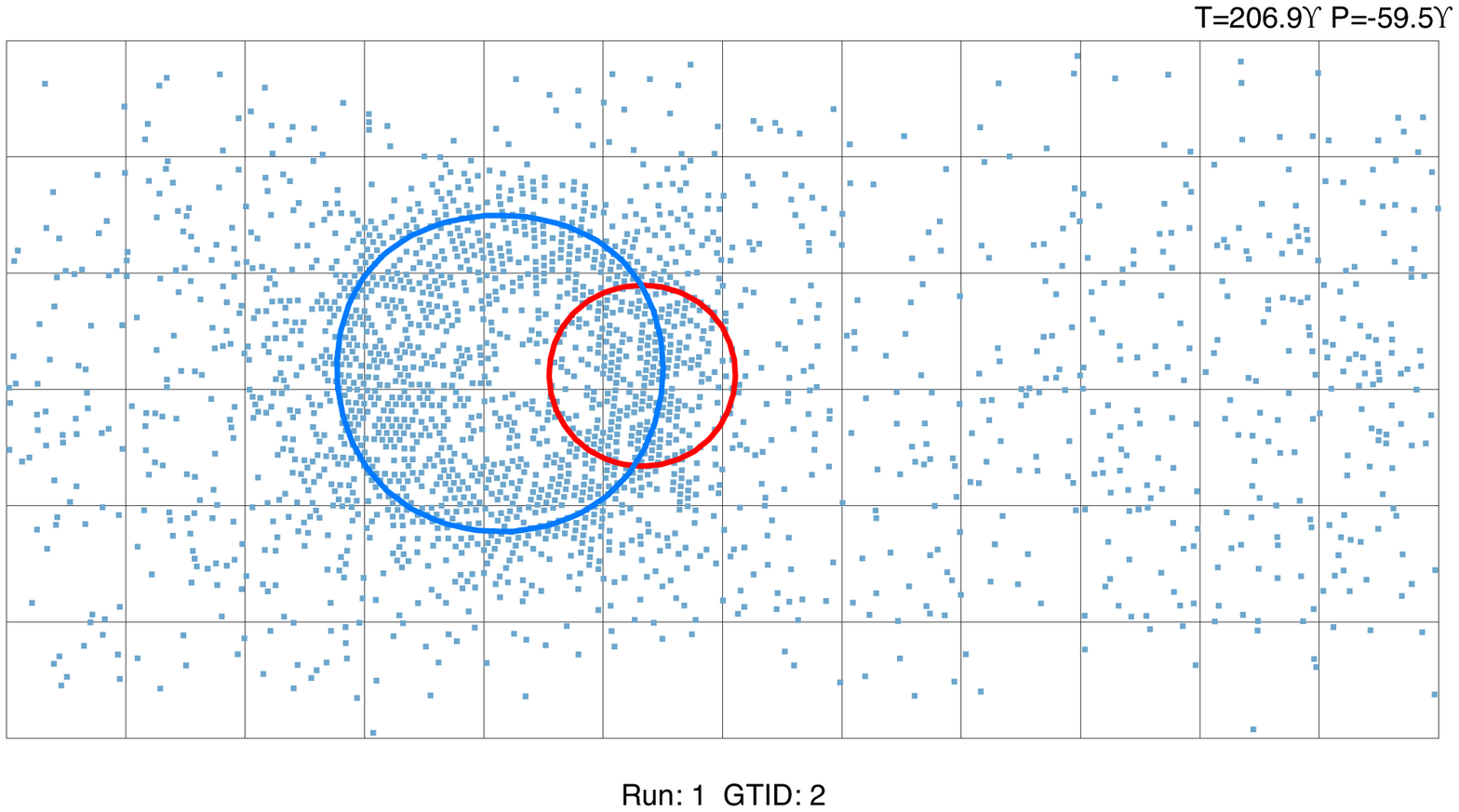}
\caption[]{Visual display of the small-ring pathology of a simulated $e^{-}$ event.  The blue ring is the primary ring, ring with the best reconstruction (i.e. minimum of  $-2\ln\lambda_{\alpha}$ in Eqn.~\ref{logLambda}), while the red is the secondary ring. This secondary ring is a MRF flaw in handling low statistics of triggered PMTs  if the local vertex ($x'_{rec}\hat{u}_{rec}$) is near the acrylic vessel. This pathology appears when the reconstruction of the local vertex from the origin is  $x'_{rec}> 350$ cm along the track.}
\label{small-ringPathology}
\end{figure}
}

\newcommand{\RingAndPoverEFigures}{
\begin{figure}[t]
 \centering
{\includegraphics[width=0.48\textwidth]{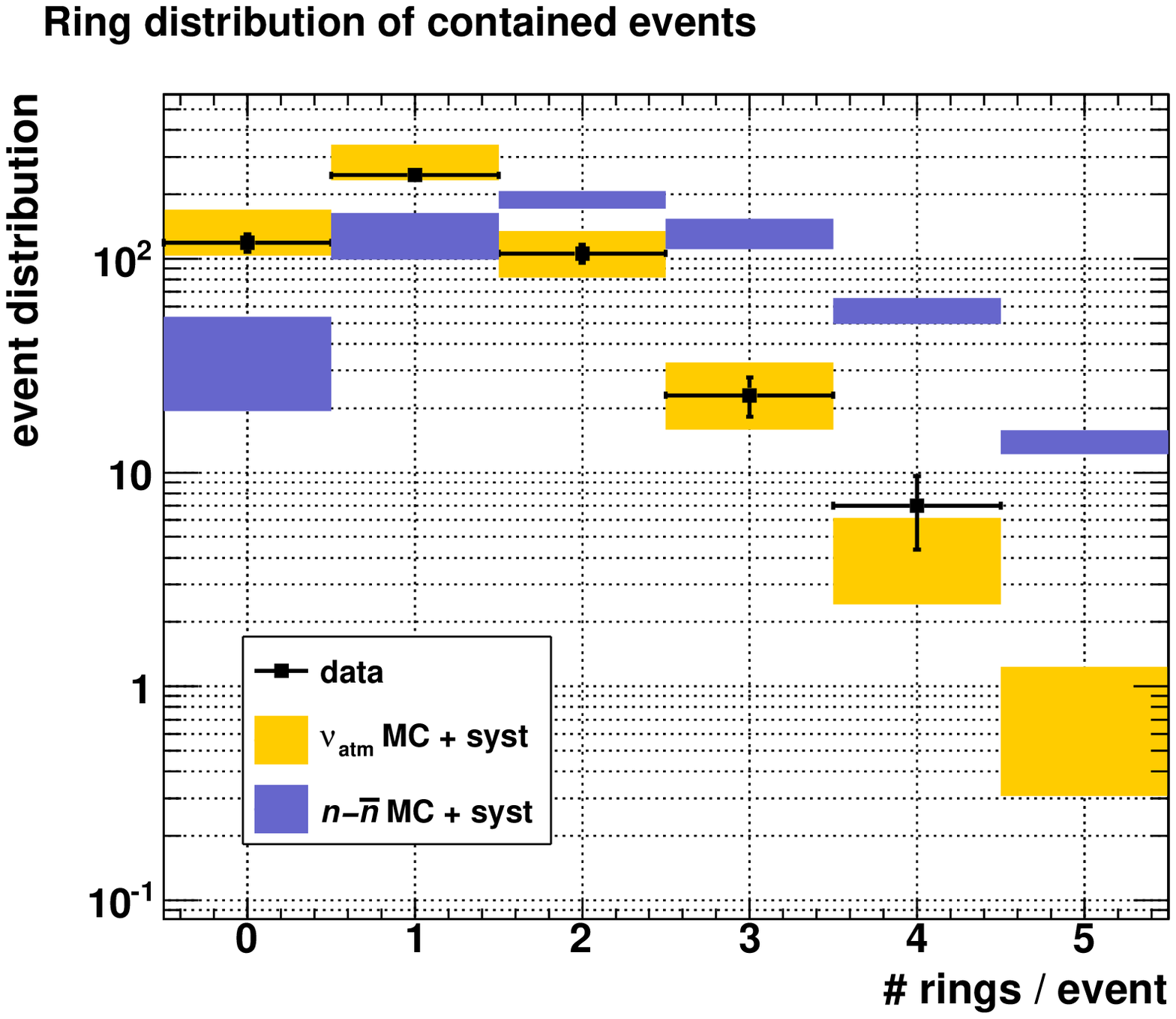}}\caption[]{{Comparison of $N_{\rm rings}$ distribution for SNO's three-phase data to simulated atmospheric neutrino and $n-\bar{n}$ oscillation events. The error bars signify all systematic uncertainties that are described in Sec.~\ref{sec:SystSect}.}}
\label{MultRing}
\end{figure}
\begin{figure}[t]
 \centering
 \includegraphics[width=0.48\textwidth]{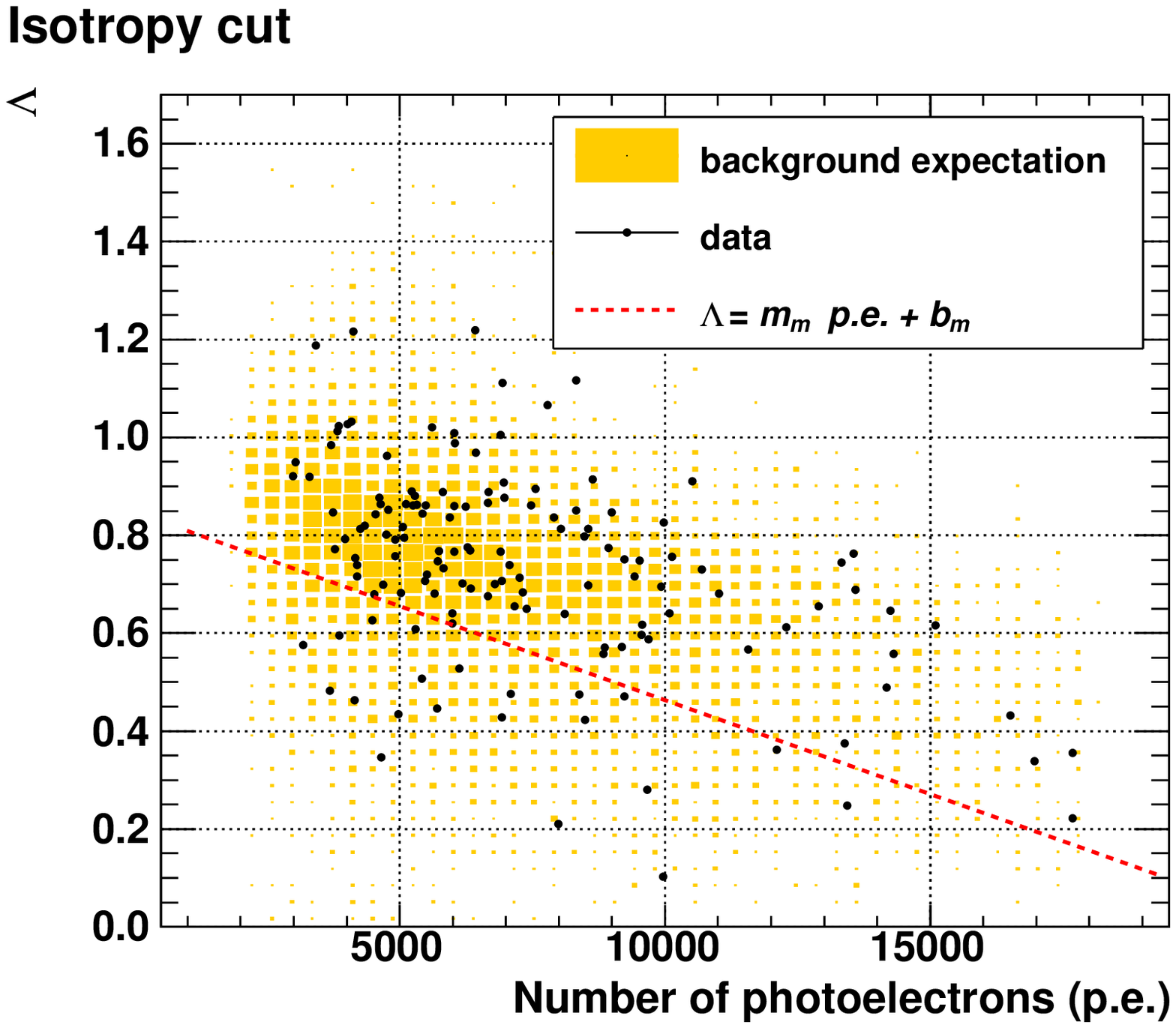}
 \includegraphics[width=0.48\textwidth]{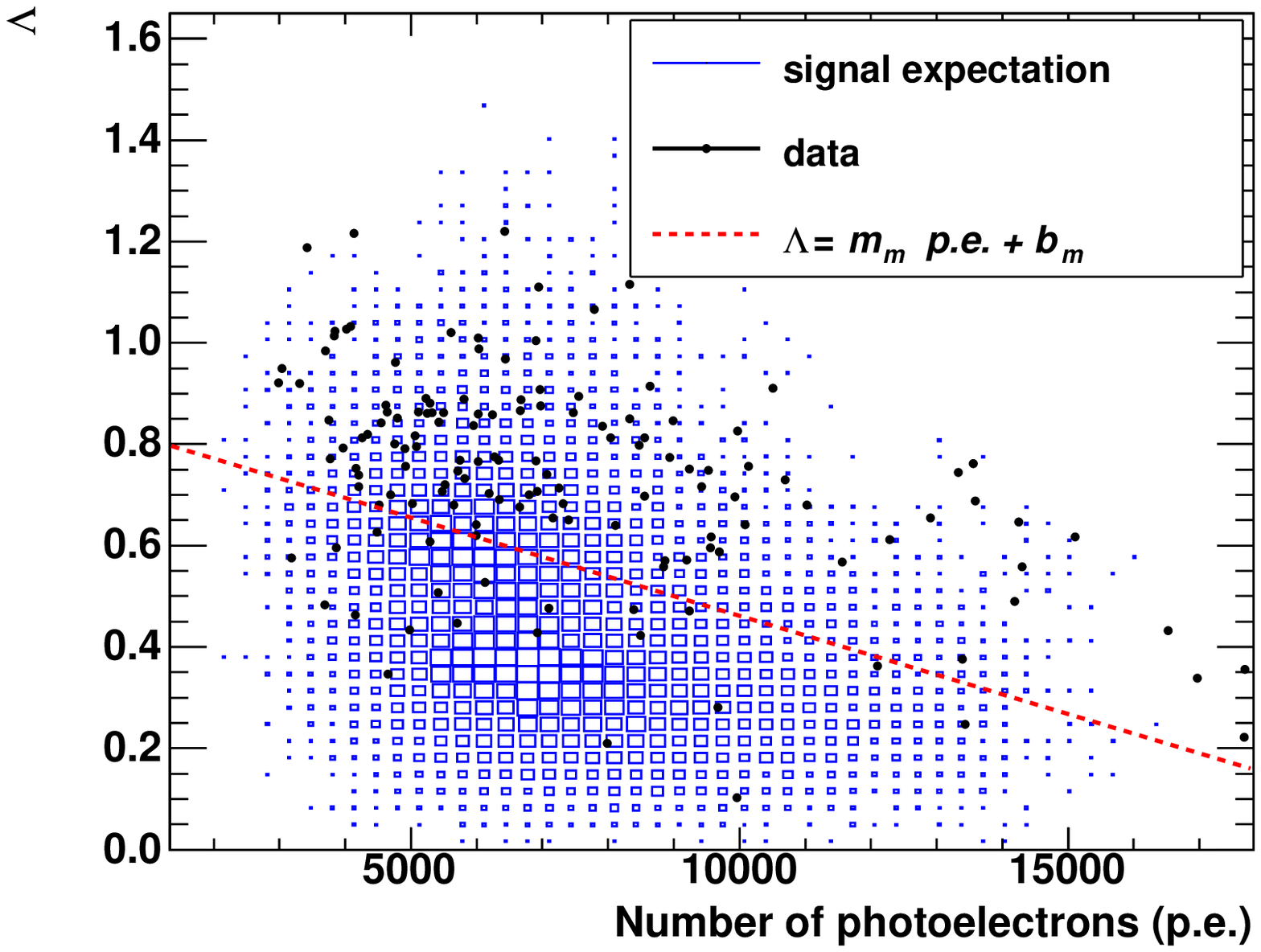}
\caption[]{Scatter plot of the $\Lambda$ parameter (Eqn.~\ref{Lambda}) and the number of photoelectrons for the background and signal expectation of multiple-ring events.  The red line denotes the isotropy cut applied to the data set to isolate signal events from background events (detailed in Sec.~\ref{event:Selection}).}
\label{3PhaseIso}
\end{figure}

}

\newcommand{\EffAndSysTable}{%
\begin{table}
{
\scriptsize
\caption[]{The $n-\bar{n}$ signal detection efficiency for momentum regimes I and II. The total efficiency for each model (shown in bold) is evaluated with the average of the efficiencies weighted by the channel branching ratios. The weighted average of the efficiency for the two momentum regimes is ($54.0\pm4.6$)\%.}
\begin{tabular}{ l l c c c}
\hline\hline
\multicolumn{5}{l}{{\bf Neutron antineutron Regime I}}\bstb\\\hline
\mcc{1}{Final State }&\mcc{1}{Channels}&\mcc{1}{$\epsilon_{\rm multi-ring}$}&\mcc{1}{$\epsilon_{\rm isotropy\: cut}$}&\mcc{1}{$\epsilon_{\rm tot}$}\bstb \\    \hline
$2\pi^{-}\pi^{+}$&$\pi^-\rho^o$              & 0.740$\pm$0.016 &0.631$\pm$0.020  & 0.467$\pm$0.018 \bst\\
&$\pi^-f^o$                                  & 0.742$\pm$0.016 &0.705$\pm$0.019  & 0.524$\pm$0.018 \\
&$\pi^-f'_2$                                 & 0.685$\pm$0.017 &0.609$\pm$0.021 & 0.417$\pm$0.018 \\
$2\pi^{-}\pi^{+}\pi^{o}$&$\pi^-\omega$  & 0.786$\pm$0.020 & 0.617$\pm$0.026 &0.485$\pm$0.024 \\
&$\pi^-X^o$							 & 0.788$\pm$0.009 & 0.721$\pm$0.011& 0.568$\pm$0.011 \\
&$\pi^-X'^{o}$							 & 0.718$\pm$0.016 & 0.631$\pm$0.021 & 0.452$\pm$0.018 \\
&$\pi^-A^{o}_2$					 & 0.756$\pm$0.015 & 0.707$\pm$0.018 & 0.534$\pm$0.018\\
&$\pi^oA^{-}_2$							 & 0.652$\pm$0.010 & 0.707$\pm$0.012 & 0.461$\pm$0.011 \\
&$\rho^-\rho^o$														 & 0.777$\pm$0.010 & 0.720$\pm$0.012 & 0.559$\pm$0.012 \\
$3\pi^{+}2\pi^{+}$  &     								     & 0.626$\pm$0.041 & 0.806$\pm$0.051 & { 0.505$\pm$0.034} \bsb\\
Total 					&&  0.734$\pm$ 0.119  &  0.719$\pm$0.117 & {\bf 0.526$\pm$0.087} \bstb\\
\hline\hline
\multicolumn{5}{l}{{\bf Neutron antineutron Regime II}}\bstb\\\hline
{Channels  } & {$\Gamma_i/\Gamma_t$} & \mcc{1}{$\epsilon_{\rm multi-ring}$}&\mcc{1}{$\epsilon_{\rm isotropy\: cut}$}&\mcc{1}{$\epsilon_{\rm tot}$} \bstb \\    \hline
$\pi^{+}\pi^{0}$ & 0.009 & 0.580$\pm$0.032 & 0.575$\pm$0.042 & 0.333$\pm$0.031 \\
$\pi^{+}2\pi^{0}$ & 0.083& 0.701$\pm$0.010 & 0.672$\pm$0.012 & 0.471$\pm$0.011 \\
$\pi^{+}3\pi^{0}$ & 0.105& 0.684$\pm$0.009 & 0.755$\pm$0.010 & 0.516$\pm$0.010 \\
$2\pi^{+}\pi^{-}\pi^{0}$& 0.223 & 0.701$\pm$0.006 & 0.748$\pm$0.007 & 0.525$\pm$0.007 \\
$2\pi^{+}\pi^{-}2\pi^{0}$ &  0.371 & 0.784$\pm$0.004 & 0.785$\pm$0.005 & 0.615$\pm$0.005 \\
$2\pi^{+}\pi^{-}\omega^{0}$ & 0.145& 0.583$\pm$0.008 & 0.829$\pm$0.008 & 0.483$\pm$0.008 \\
$3\pi^{+}2\pi^{-}\pi^{0}$ &0.064 & 0.545$\pm$0.013 & 0.809$\pm$0.013 & 0.441$\pm$0.012 \\
Total & 1.000& 0.702$\pm$0.003 & 0.769$\pm$0.003 & {\bf 0.540$\pm$0.003 }\bstb\\\hline\hline\end{tabular}
\label{nnbarChan}
}
\end{table}
\begin{table}
{
\scriptsize
\caption[]{Atmospheric neutrino detection efficiency. The total efficiency for each model (shown in bold) is evaluated with the average of the efficiencies weighted by the channel branching ratios. The application of the multiple-ring and isotropy cuts resulted in the rejection of 94.6\% of the contained atmospheric neutrino events (Eqn.~\ref{eq:atmointeractions}) over the three SNO phases.}
\begin{tabular}{ l c c c c}
\hline\hline
\multicolumn{5}{l}{{\bf Atmospheric neutrino MC efficiencies}}\bstb\\\hline
\mcc{1}{$\Gamma_i$  } & \mcc{1}{$\Gamma_i/\Gamma_t$} & \mcc{1}{$\epsilon_{\rm multi-ring}$}&\mcc{1}{$\epsilon_{\rm isotropy\: cut}$}&\mcc{1}{$\epsilon_{\rm tot}$} \bstb \\    \hline
\hline\hline\multicolumn{5}{l}{{Phase-I}}\bstb\\\hline

$\nu_{\rm cc}$ & 0.476$\pm$0.005 & 0.180$\pm$0.006 & 0.102$\pm$0.011 & 0.018$\pm$0.002 \\
$\nu_{\rm nc}$ & 0.047$\pm$0.002 & 0.300$\pm$0.022 & 0.256$\pm$0.038 & 0.077$\pm$0.013 \\
$\nu_{\pi}$ & 0.372$\pm$0.005 & 0.296$\pm$0.008 & 0.250$\pm$0.014 & 0.074$\pm$0.004 \\
$\nu_{n\pi}$ & 0.103$\pm$0.003 & 0.356$\pm$0.015 & 0.289$\pm$0.024 & 0.103$\pm$0.010 \\
$\nu_{\rm otr}$ & 0.001$\pm$0.000 & 0.417$\pm$0.128 & 0.000$\pm$0.124 & 0.000$\pm$0.066 \\
$\nu_{\rm tot}$ & 1.000$\pm$0.000 & 0.247$\pm$0.004 & 0.204$\pm$0.008 &  {\bf 0.051$\pm$0.002} \\

\hline\hline\multicolumn{5}{l}{{Phase-II}}\bstb\\\hline
$\nu_{\rm cc}$ & 0.470$\pm$0.004 & 0.192$\pm$0.005 & 0.128$\pm$0.010 & 0.025$\pm$0.002 \\
$\nu_{\rm nc}$ & 0.054$\pm$0.002 & 0.351$\pm$0.018 & 0.310$\pm$0.029 & 0.109$\pm$0.012 \\
$\nu_{\pi}$ & 0.365$\pm$0.004 & 0.283$\pm$0.006 & 0.227$\pm$0.011 & 0.064$\pm$0.004 \\
$\nu_{n\pi}$ & 0.110$\pm$0.003 & 0.366$\pm$0.013 & 0.306$\pm$0.020 & 0.112$\pm$0.008 \\
$\nu_{\rm otr}$ & 0.001$\pm$0.000 & 0.467$\pm$0.118 & 0.429$\pm$0.157 & 0.200$\pm$0.100 \\
$\nu_{\rm tot}$ & 1.000$\pm$0.000 & 0.253$\pm$0.004 & 0.211$\pm$0.007 &  {\bf 0.053$\pm$0.002} \\

\hline\hline\multicolumn{5}{l}{{Phase-III}}\bstb\\\hline
$\nu_{\rm cc}$ & 0.457$\pm$0.005 & 0.182$\pm$0.006 & 0.170$\pm$0.013 & 0.031$\pm$0.003 \\
$\nu_{\rm nc}$ & 0.059$\pm$0.002 & 0.346$\pm$0.019 & 0.434$\pm$0.034 & 0.150$\pm$0.014 \\
$\nu_{\pi}$ & 0.370$\pm$0.005 & 0.251$\pm$0.007 & 0.258$\pm$0.014 & 0.065$\pm$0.004 \\
$\nu_{n\pi}$ & 0.113$\pm$0.003 & 0.324$\pm$0.014 & 0.315$\pm$0.024 & 0.102$\pm$0.009 \\
$\nu_{\rm otr}$ & 0.001$\pm$0.000 & 0.286$\pm$0.149 & 0.500$\pm$0.224 & 0.143$\pm$0.131 \\
$\nu_{\rm tot}$ & 1.000$\pm$0.000 & 0.233$\pm$0.004 & 0.251$\pm$0.009 & {\bf 0.059$\pm$0.002} \\

\hline\hline
\end{tabular}
\label{atmoChan}
}
\end{table}
\begin{table*}
\centering
{
\caption[Systematic uncertainties of $n-\bar{n}$ and atmospheric efficiency.]{Systematic uncertainties of $n-\bar{n}$ detection efficiency ($\epsilon^{n-\bar{n}}$), the atmospheric neutrino detection efficiency ($\epsilon^{\rm atmo}$) and the atmospheric neutrino background rate ($\Phi^{\rm atmo}$) for phases I, II, and III. The overall signal detection efficiency uncertainty is 11.7\%, while the overall background systematic uncertainty is 24.5\%.}
\begin{tabular}{ c  c   r c l  r c l  r c l }
\hline\hline
\mcc{1}{ } & \mcl{1}{Phase independent}&\mcc{3}{Phase I}&\mcc{3}{Phase II}& \mcc{3}{Phase III}\\
{Uncertainty}&$\Phi_{\rm osc}^{\rm atmo}$&$\Phi_{\rm I}^{\rm atmo}$&$\epsilon_{\rm I}^{\rm atmo}$&$\epsilon_{\rm I}^{n-\bar{n}}$  &$\Phi_{\rm II}^{\rm atmo}$&$\epsilon_{\rm II}^{\rm atmo}$&$\epsilon_{\rm II}^{n-\bar{n}}$  &$\Phi_{\rm III}^{\rm atmo}$ & $\epsilon_{\rm III}^{\rm atmo}$&$\epsilon_{\rm III}^{n-\bar{n}}$ \bstb \\    \hline\hline
\multicolumn{5}{l}{{{\it Measurement uncertainties}}}\bst\\\hline
photoelectrons             				& |  & 1.5\%  & 0.2\%  &  0.1\%   & 1.8\%  & 0.3\%     & 0.1\% 	 & 1.7\% & 0.3\%  & 0.1\% 	  \bst \\
MRF calibration  				& |  & |	& 16.7\% & 6.5\%    & |    & 14.4\%    & 5.6\%     & |   & 15.4\% & 8.3\%     \\
$\cos\theta_{\rm ring}$				& |  & |	& 8.4\%  & 1.4\%    & |    & 8.1\%     & 1.6\%     & |   & 9.2\%  & 2.2\%  \bsb\\
\hline\multicolumn{5}{l}{{{\it Model uncertainties}}}\bst\\\hline
 $\nu_{\rm atmo}$ models		& |  & |    & 6.5\%  & |      & |    & 6.5\%     & |       & |   & 6.5\%  & |    \\
$\bar{n}p$ modeling			    & |  & |    & |    & 9.4\%  	& |    & |       & 9.4\%     & |   & |    & 9.4\%   \bsb\\
\hline\multicolumn{5}{l}{{{\it External input uncertainties}}}\bst\\\hline
$\phi_{\rm normalization}$ (SNO)	& 7.4\%& |  & |    & |      & |  	 & |       & |       & |   & |   & | \\
$\Delta m^2_{\rm MINOS}$ 			&$<$0.01\%& |    & |    & |      & |  	 & |       & |       & |   & |   & | 		\\
$\sin^{2}2\theta_{\rm SK}$ 			& 0.7\%  & |    & |    & |      & |  	 & |       & |       & |   & |   & | \\
$\Delta$ Resonance (20\%)   	& |  & 8.0\%  & 10.6\% & |      & 8.4\%  & 11.5\%    & |       & 8.5\% & 10.9\% & |    \\
$\bar{\nu}/\nu$ ratio (SK)			& |  & 1.4\%  & 1.4\%  & |    & 1.5\%     & 1.5\%   & |	    & 1.5\% & 1.5\%  & |      \bsb\\
\hline\hline
{\bf Total} 			 		& 7.4\%  & 8.3\%  & 22.5\% & 11.5\%   & 8.7\%  & 21.2\%    & 11.1\%    & 8.8\% & 22.0\% & 12.7\% \bstb\\
\hline\hline
\label{table:SystErr}
\end{tabular}
}
\end{table*}

}

\newcommand{\LimitResultTable}{%
\begin{table*}
\centering
{\small
\caption[]{$T_{\rm intranuclear}$ and $\tau_{\rm free}$ limit obtain for the SNO experiment evaluated with the profile likelihood method.  The bounded/unbounded (B/UB) limits for the free oscillation time are listed in the right-most column. }
\label{nnbarLimit2}
\begin{tabular}{   l  c  c  c  c  c  c  c  c   c  c c}
\hline\hline
SNO Phase  &{Exposure}     & $\epsilon_{\rm tot}$ & Observed & Bkgd & UL & $T_{\rm intranuclear}$ & R &$\tau_{\rm free}$\bst \\
     				    	&$10^{31}$ n-yr&         (\%)	&		& 		&  (B/UB)	&	$10^{31}$ yr	& $10^{23}$ s$^{-1}$ &10$^8$ s (B/UB)\bsb \\    \hline\hline
Phase I    & 5.78 & 55.0 & 8 &  7.8 & ( 6.66 /  6.66) & ( 0.48 /  0.48) & 0.248 &  (0.78 / 0.78)\\
Phase II    & 8.23 & 55.8 & 10 & 12.2 & ( 5.91 /  5.52) & ( 0.78 /  0.83) & 0.248 &  (1.00 / 1.03)\\
Phase III    & 6.47 & 50.9 & 5 & 10.5 & ( 3.09 /  0.63) & ( 1.06 /  5.25) & 0.248 &  (1.16 / 2.59)\\
Combined Phases & 20.47 & 54.0 & 23 & 30.5 & ( 9.38 /  7.46) & ( 1.18 /  1.48) & 0.248 &  (1.23 / 1.37)\\
\hline\hline
\end{tabular}
}
\end{table*}
}

\newcommand{\resultTable}{
\begin{table}[t]
{
\small
\caption[]{The number of contained events in data ($x$) versus the expected background ($b$) at each stage of the event selection. The criteria are the contained event (cont), the Multiple Ring (MR) and the Isotropy Cut (IC) criteria as explained in Sec.~\ref{event:Selection}.  The IC criteria is the last of the analysis and these values represent the measured signal ($x_{IC}$) and expected background ($b_{ic}$) for this analysis.}
\begin{tabular}{ l r r r r rr  r} \hline\hline
Phase & $x_{cont}$ & $b_{cont}$ & $x_{MR}$& $b_{MR}$ &  &$x_{IC}$ & $b_{IC}$ \bstb\\ \hline
Phase I    &  143  & 154.1   &  43     & 38.1   & &  {\bf 8}       &    {\bf7.8}       \bst\\
Phase II   &  188  & 228.2   &  54     & 57.8    &&  {\bf10}     &  {\bf12.2}     \\
Phase III  &  170  & 179.4   &  39     & 41.9   & &   {\bf 5}    &   {\bf10.5}    \bsb\\\hline
total          &  501  &  561.7  &  136   & 137.8 & &  {\bf23 }  &  {\bf30.5}    \bstb \\
\hline\hline
\end{tabular}
\label{3PhasesResults}
}
\end{table}
}

\begin{document}

\title{The search for neutron-antineutron oscillations at the Sudbury Neutrino Observatory}

\newcommand{\dec}{Deceased}
\newcommand{\alta}{Department of Physics, University of
Alberta, Edmonton, Alberta, T6G 2R3, Canada}
\newcommand{\ubc}{Department of Physics and Astronomy, University of
British Columbia, Vancouver, BC V6T 1Z1, Canada}
\newcommand{\bnl}{Chemistry Department, Brookhaven National
Laboratory,  Upton, NY 11973-5000}
\newcommand{\carleton}{Ottawa-Carleton Institute for Physics, Department of Physics, Carleton University, Ottawa, Ontario K1S 5B6, Canada}
\newcommand{\carletona}{Department of Physics, Carleton University, Ottawa, Ontario, Canada}
\newcommand{\uog}{Physics Department, University of Guelph,
Guelph, Ontario N1G 2W1, Canada}
\newcommand{\lu}{Department of Physics and Astronomy, Laurentian
University, Sudbury, Ontario P3E 2C6, Canada}
\newcommand{\lbnl}{Institute for Nuclear and Particle Astrophysics and
Nuclear Science Division, Lawrence Berkeley National Laboratory, Berkeley, CA 94720-8153}
\newcommand{\lbla}{ Lawrence Berkeley National Laboratory, Berkeley, CA}
\newcommand{\lanl}{Los Alamos National Laboratory, Los Alamos, NM 87545}
\newcommand{\llnl}{Lawrence Livermore National Laboratory, Livermore, CA. Supported under the auspices of the U.S. Department of Energy by Lawrence Livermore National Laboratory under Contract DE-AC52-07NA27344, release no. LLNL-JRNL-728820}
\newcommand{\lanla}{Los Alamos National Laboratory, Los Alamos, NM 87545}
\newcommand{\oxford}{Department of Physics, University of Oxford,
Denys Wilkinson Building, Keble Road, Oxford OX1 3RH, UK}
\newcommand{\penn}{Department of Physics and Astronomy, University of
Pennsylvania, Philadelphia, PA 19104-6396}
\newcommand{\pennx}{Department of Physics and Astronomy, University of
Pennsylvania, Philadelphia, PA}
\newcommand{\queens}{Department of Physics, Queen's University,
Kingston, Ontario K7L 3N6, Canada}
\newcommand{\uw}{Center for Experimental Nuclear Physics and Astrophysics,
and Department of Physics, University of Washington, Seattle, WA 98195}
\newcommand{\uwx}{Center for Experimental Nuclear Physics and Astrophysics,
and Department of Physics, University of Washington, Seattle, WA}
\newcommand{\uta}{Department of Physics, University of Texas at Austin, Austin, TX 78712-0264}
\newcommand{\triumf}{TRIUMF, 4004 Wesbrook Mall, Vancouver, BC V6T 2A3, Canada}
\newcommand{\ralimp}{Rutherford Appleton Laboratory, Chilton, Didcot, UK} 
\newcommand{\iusb}{Department of Physics and Astronomy, Indiana University, South Bend, IN}
\newcommand{\fnal}{Fermilab, Batavia, IL}
\newcommand{\uo}{Department of Physics and Astronomy, University of Oregon, Eugene, OR}
\newcommand{\hu}{Department of Physics, Hiroshima University, Hiroshima, Japan}
\newcommand{\slac}{Stanford Linear Accelerator Center, Menlo Park, CA}
\newcommand{\mac}{Department of Physics, McMaster University, Hamilton, ON}
\newcommand{\doe}{US Department of Energy, Germantown, MD}
\newcommand{\lund}{Department of Physics, Lund University, Lund, Sweden}
\newcommand{\mpi}{Max-Planck-Institut for Nuclear Physics, Heidelberg, Germany}
\newcommand{\uom}{Ren\'{e} J.A. L\'{e}vesque Laboratory, Universit\'{e} de Montr\'{e}al, Montreal, PQ}
\newcommand{\cwru}{Department of Physics, Case Western Reserve University, Cleveland, OH}
\newcommand{\pnnl}{Pacific Northwest National Laboratory, Richland, WA}
\newcommand{\uc}{Department of Physics, University of Chicago, Chicago, IL}
\newcommand{\mitt}{Laboratory for Nuclear Science, Massachusetts Institute of Technology, Cambridge, MA 02139}
\newcommand{\ucsd}{Department of Physics, University of California at San Diego, La Jolla, CA }
\newcommand{	\lsu	}{Department of Physics and Astronomy, Louisiana State University, Baton Rouge, LA 70803}
\newcommand{\imp}{Imperial College, London, UK}
\newcommand{\uci}{Department of Physics, University of California, Irvine, CA 92717}
\newcommand{\ucia}{Department of Physics, University of California, Irvine, CA}
\newcommand{\suss}{Department of Physics and Astronomy, University of Sussex, Brighton  BN1 9QH, UK}
\newcommand{\sussx}{Department of Physics and Astronomy, University of Sussex, Brighton, UK}
\newcommand{\lifep}{Laborat\'{o}rio de Instrumenta\c{c}\~{a}o e F\'{\i}sica Experimental de
Part\'{\i}culas, Av. Elias Garcia 14, 1$^{\circ}$, 1000-149 Lisboa, Portugal}
\newcommand{\lipx}{Laborat\'{o}rio de Instrumenta\c{c}\~{a}o e F\'{\i}sica Experimental de
Part\'{\i}culas,  Lisboa, Portugal}
\newcommand{\hku}{Department of Physics, The University of Hong Kong, Hong Kong.}
\newcommand{\aecl}{Atomic Energy of Canada, Limited, Chalk River Laboratories, Chalk River, ON K0J 1J0, Canada}
\newcommand{\nrc}{National Research Council of Canada, Ottawa, ON K1A 0R6, Canada}
\newcommand{\princeton}{Department of Physics, Princeton University, Princeton, NJ 08544}
\newcommand{\birkbeck}{Birkbeck College, University of London, Malet Road, London WC1E 7HX, UK}
\newcommand{\snoi}{SNOLAB, Lively, ON P3Y 1N2, Canada}
\newcommand{\uba}{University of Buenos Aires, Argentina}
\newcommand{\hvd}{Department of Physics, Harvard University, Cambridge, MA}
\newcommand{\pny}{Goldman Sachs, 85 Broad Street, New York, NY}
\newcommand{\pnv}{Remote Sensing Lab, PO Box 98521, Las Vegas, NV 89193}
\newcommand{\nts}{Nevada National Security Site, Las Vegas, NV}
\newcommand{\psis}{Paul Schiffer Institute, Villigen, Switzerland}
\newcommand{\liverpool}{Department of Physics, University of Liverpool, Liverpool, UK}
\newcommand{\uto}{Department of Physics, University of Toronto, Toronto, ON, Canada}
\newcommand{\uwisc}{Department of Physics, University of Wisconsin, Madison, WI}
\newcommand{\psu}{Department of Physics, Pennsylvania State University,
     University Park, PA}
\newcommand{\anl}{Deparment of Mathematics and Computer Science, Argonne
     National Laboratory, Lemont, IL}
\newcommand{\cornell}{Department of Physics, Cornell University, Ithaca, NY}
\newcommand{\tufts}{Department of Physics and Astronomy, Tufts University, Medford, MA}
\newcommand{\ucd}{Department of Physics, University of California, Davis, CA}
\newcommand{\unc}{Department of Physics, University of North Carolina, Chapel Hill, NC}
\newcommand{\dresden}{Institut f\"{u}r Kern- und Teilchenphysik, Technische Universit\"{a}t Dresden,  Dresden, Germany} 
\newcommand{\isu}{Department of Physics, Idaho State University, Pocatello, ID}
\newcommand{\qmul}{Dept. of Physics, Queen Mary University, London, UK}
\newcommand{\ucsb}{Dept. of Physics, University of California, Santa Barbara, CA}
\newcommand{\cern}{CERN, Geneva, Switzerland}
\newcommand{\utah}{Dept. of Physics, University of Utah, Salt Lake City, UT}
\newcommand{\casa}{Center for Astrophysics and Space Astronomy, University
of Colorado, Boulder, CO}
\newcommand{\susel}{Sanford Underground Research Laboratory, Lead, SD}  
\newcommand{\ntu}{Center of Cosmology and Particle Astrophysics, National Taiwan University, Taiwan}
\newcommand{\berlin}{Institute for Space Sciences, Freie Universit\"{a}t Berlin,
Leibniz-Institute of Freshwater Ecology and Inland Fisheries, Germany}
\newcommand{\bhsu}{Black Hills State University, Spearfish, SD} 
\newcommand{\queensa}{Dept.\,of Physics, Queen's University,
Kingston, Ontario, Canada} 
\newcommand{\aasu}{Dept.\,of Chemistry and Physics, Armstrong  State University, Savannah, GA}
\newcommand{\ucb}{Physics Department, University of California at Berkeley, Berkeley, CA 94720-7300}
\newcommand{\ucbx}{Physics Department, University of California at Berkeley, and Lawrence Berkeley National Laboratory, Berkeley, CA}
\newcommand{\mcgill}{Physics Department, McGill University, Montreal, QC, Canada}
\newcommand{\columbia}{Columbia University, New York, NY}
\newcommand{\rhul}{Dept. of Physics, Royal Holloway University of London, Egham, Surrey, UK}
\newcommand{\ubama}{Department of Physics and Astronomy, University of Alabama, Tuscaloosa, AL}
\newcommand{\kit}{Institut f\"{u}r Experimentelle Kernphysik, Karlsruher Institut f\"{u}r Technologie, Karlsruhe, Germany}
\newcommand{\winnipeg}{Department of Physics, University of Winnipeg, Winnipeg, Manitoba, Canada}
\newcommand{\kwantlen}{Kwantlen Polytechnic University, Surrey, BC, Canada}
\newcommand{\cea}{CEA-Saclay, DSM/IRFU/SPP, Gif-sur-Yvette, France}
\newcommand{\sunysb}{Laufer Center, Stony Brook University, Stony Brook, NY}
\newcommand{\rock}{Rock Creek Group, Washington, DC}
\newcommand{\rcnp}{Research Center for Nuclear Physics, Osaka, Japan}
\newcommand{\usd}{University of South Dakota, Vermillion, SD}
\newcommand{\lancaster}{Physics Department, Lancaster University, Lancaster, UK}
\newcommand{\potsdam}{GFZ German Research Centre for Geosciences, Potsdam, Germany}
\newcommand{\kirchhoff}{Ruprecht-Karls-Universit\"{a}t Heidelberg, Im Neuenheimer Feld 227, Heidelberg, Germany}
\newcommand{\continuum}{Continuum Analytics,  Austin, TX}
\newcommand{\usaid}{Global Development Lab, U.S. Agency for International Development, Washington DC}


\affiliation{\alta}
\affiliation{\ucb}
\affiliation{\ubc}
\affiliation{\bnl}
\affiliation{\carleton}
\affiliation{\uog}
\affiliation{\lu}
\affiliation{\lbnl}
\affiliation{\lifep}
\affiliation{\lanl}
\affiliation{\lsu}
\affiliation{\mitt}
\affiliation{\oxford}
\affiliation{\penn}
\affiliation{\queens}
\affiliation{\snoi}
\affiliation{\uta}
\affiliation{\triumf}
\affiliation{\uw}
\author{B.~Aharmim}\affiliation{\lu}
\author{S.\,N.~Ahmed}\affiliation{\queens}
\author{A.\,E.~Anthony}\altaffiliation{Present address: \usaid}\affiliation{\uta}
\author{N.~Barros}\altaffiliation{Present address: \pennx}\affiliation{\lifep}
\author{E.\,W.~Beier}\affiliation{\penn}
\author{A.~Bellerive}\affiliation{\carleton}
\author{B.~Beltran}\affiliation{\alta}
\author{M.~Bergevin}\altaffiliation{Present address: \llnl}\affiliation{\lbnl}\affiliation{\uog}
\author{S.\,D.~Biller}\affiliation{\oxford}
\author{K.~Boudjemline}\affiliation{\carleton}\affiliation{\queens}
% ...

\collaboration{SNO Collaboration}
\noaffiliation

\date{\today}

\begin{abstract}
Tests on $B-L$ symmetry breaking models are important probes to search for new physics. One proposed model with $\Delta(B-L)=2$ involves the oscillations of a neutron to an antineutron. In this paper a new limit on this process is derived for the data acquired from all three operational phases of the Sudbury Neutrino Observatory  experiment. The search was concentrated in oscillations occurring within the deuteron, and 23 events are observed against a background expectation of 30.5 events. These translate to a lower limit on the nuclear lifetime of $1.48\times 10^{31}$ years at 90\% confidence level (CL) when no restriction is placed on the signal likelihood space (unbounded). Alternatively, a lower limit on the nuclear lifetime was found to be $1.18\times 10^{31}$ years at 90\% CL when the signal was forced into a positive likelihood space (bounded). Values for the free oscillation time derived from various models are also provided in this article. This is the first search for neutron-antineutron oscillation with the deuteron as a target.
\end{abstract}


\maketitle


\section{Introduction}

One of the Sakharov conditions dictates that the baryon number $B$ must be violated in order to obtain the imbalance between matter and antimatter seen in the universe today \cite{Sakharov}.  Proton decay is one example of a process that would violate $B$; however, this process has not yet been observed and the current experimental limits exceed the early theoretical estimates by orders of magnitude.  Traditional proton decay modes rely not only on $B$ violation, but also on lepton number ($L$) violation. This is possible if an underlying $B-L$ quantum number exists and is conserved via a U$(1)_{B-L}$ gauge group.  In this article an experimental limit on neutron-antineutron oscillation, a process that violates purely the quantum number $B$ is presented.  In a U$(1)_{B-L}$ gauge group, a violation of $B$ also results in a violation of $B-L$.

As an example, the proton decay mode $p\rightarrow e^+\pi^0$ has a baryon number change of $\Delta(B)=1$, a lepton number change of $\Delta(L)=-1$, resulting in the $B-L$ quantum number being conserved.  
In comparison, a neutron transforming into an antineutron $n\rightarrow \bar{n}$ is a process that violates the $B$ quantum number, and by construct the $B-L$ quantum number, by two.  
However, if $B-L$ is not a GUT symmetry and $L$ is a conserved quantity, then the $n\rightarrow \bar{n}$ process provides a mechanism that involves solely $B$ violation \cite{MohapatraReview}.  A discovery of this process would bridge the gap in our understanding of the matter-antimatter asymmetry.

An in-depth experimental and theoretical review of neutron-antineutron physics and baryon and lepton number violation can be found in \cite{nnbarReview}.

Two experimental scenarios exist in which the neutron-antineutron oscillation process is potentially observable: (1) the oscillations of neutrons to antineutrons in bound nuclei, and (2) the oscillations of a beam of cold neutrons against an annihilation target situated at an optimized distance \cite{nnbarReview}. This paper will concentrate on the former scenario, in which the antineutrons interact with the surrounding nucleons and produce a GeV-scale signature.

The oscillation process is suppressed within the nuclear environment. The intra-nuclear (case 1) and free measurements (case 2) are related to each other by
\begin{equation}
T_{\rm intranuclear}=\tau_{\rm free}^2 R
\end{equation}
where $T_{\rm intranuclear}$ is the lifetime of a neutron in the intra-nuclear media, $\tau_{\rm free}$ is the oscillation time outside an intra-nuclear environment, and $R$ is the suppression factor which is target-dependent. In intra-nuclear experiments, the rate reduction due to the suppression factor needs to be offset by the exposure to a large quantity of bound neutrons, requiring kiloton scale experiments.

The suppression factor varies for different nuclei and can be derived from theoretical models  \cite{nnbarReview}. The magnitude of this suppression is proportional to the potential energy of the neutron inside the nucleus. Since the SNO experiment was filled with heavy water ($^2$H$_2$O, denoted as D$_2$O hereafter), the deuteron ($^2$H) is an intra-nuclear source for neutron-antineutron oscillations, and has a lower suppression factor compared to oxygen by a factor of four on average \cite{PhysRevC.79.054001}.

It is taken as a convention in this article that $\bar{n}n$ refers to the collision of $\bar{n}$ with $n$ while $n$-$\bar{n}$ refers to the GeV-scale signature of intra-nuclear neutron-antineutron oscillations. The signature for this process consists of multi-prong events of multiple charged and neutral pions from $\bar{n}p$ or $\bar{n}n$ interactions. Some of these pions can be absorbed by the $^{16}$O before leaving the nucleus, which can lead to issues with momentum and energy reconstruction due to the missing energy.

The current experimental limits are of the order  $\tau_{\rm free}\sim10^{8}$~s \cite{ILL,SoudanII,IMB,nnbarSuperK}.  Calculations using seesaw models with parity symmetry predict an upper limit to the free oscillation time of $\tau_{\rm free}=\hbar/\delta m_{n-\bar{n}}c^2<10^{10}$~s \cite{Babu2001269}, where $\delta m_{n-\bar{n}}$ is a perturbation term equivalent to the mixing rate of neutrons to antineutrons.

The most recent measurement of free neutron-antineutron oscillations set a lower $\tau_{\rm free}$  limit of 0.86$\times$10$^{8}$~s at 90\% CL \cite{ILL}. A  measurement using $^{56}$Fe was made at the Soudan II experiment of $T_{\rm intranuclear}>$7.2$\times$10$^{31}$~years at 90\% CL corresponding to a free oscillation limit of $1.3\times10^{8}$~s at 90\% CL \cite{SoudanII}. The Soudan II analysis used a multi-prong approach, requiring four distinct particle tracks and kinematic constraints to evaluate the rate of $n$-$\bar{n}$ events.

The most recent measurement of the nuclear bounded neutron-antineutron oscillations in $^{16}$O was published by the Super-Kamiokande experiment, which set a limit of $T_{\rm intranuclear}>19\times10^{31}$~years at 90\% CL corresponding to a free oscillation limit of $2.7\times10^{8}$~s at 90\% CL \cite{nnbarSuperK}. The Super-Kamiokande analysis required careful modeling of the effect of pion absorption in $^{16}$O in the multiple-prong signature of  $n$-$\bar{n}$ events.

The Sudbury Neutrino Observatory (SNO) was a heavy-water Cherenkov ring-imaging detector that could search for neutron-antineutron oscillations with high sensitivity. The large deuteron abundance allows a competitive search for $n$-$\bar{n}$ in a two-nucleon system. Since no surrounding nucleons are present after an $n$-$\bar{n}$ occurs in the deuteron, no immediate pion absorption is possible leading to a higher detection efficiency. This paper presents the first search for $n$-$\bar{n}$ using the deuteron as a source.

The analysis presented in this article will also focus on a multi-prong approach with constraints on the visible energy. Prongs are identified by reconstructing the Cherenkov rings created by the charged particle tracks, and it is required that at least two separate prongs are observed. Particle identification (e.g. $e^\pm,\mu^\pm,\pi^\pm,\pi^0$,...) is not made for each prong; it was found to be adequate for this analysis to simply count the number of rings in an event. Since there is no particle identification, a parameter to evaluate the spatial isotropy of all reconstructed rings or prongs ($\Lambda$) is used in lieu of momentum reconstruction in this analysis.

This paper is organized as follows. In Sec.~\ref{Section:SNODETECTOR}, an overview of the SNO operational phases is given, and the total exposure for the neutron-antineutron oscillations search is provided.  In Sec.~\ref{Section:NNBAR}, a brief description is given of the current $n$-$\bar{n}$ theoretical models and the expected detector response for this process in both $^2$H and $^{16}$O nuclei with a focus on why $n$-$\bar{n}$ oscillation in $^2$H is studied in this paper.  Section~\ref{Section:AtmoNu} describes the backgrounds in an $n$-$\bar{n}$ search: atmospheric neutrino background and other interactions that can mimic the signal will be detailed.

Section~\ref{Section:PIDReco} details the reconstruction techniques used for signal and background characterization and explores the case of the propagation of charged pions, which are produced when an antineutron annihilates with a neighbor nucleon.  The behavior of charged pions in the heavy water in SNO is different from that in a traditional water Cherenkov detector such as Super-Kamiokande. This difference will be highlighted in this section.

Section~\ref{event:Selection} details the selection criteria put in place to distinguish the signal from the backgrounds and presents the systematic uncertainties in this analysis.  In Sec.~\ref{Section:Results}, the technique used to evaluate the limit on the neutron-antineutron signal from the observed events is explained and the results of the analysis are presented.

\section{The SNO Detector}\label{Section:SNODETECTOR}

The SNO detector was a heavy-water Cherenkov imaging detector located at a depth of 2.092 km (5890 $\pm$ 94 meters water equivalent) in INCO's (now VALE's) Creighton \#9 nickel mine near Sudbury, Ontario, Canada. The experiment took data between November 2, 1999 and November 28, 2006. It consisted of 1000 metric tons (tonnes) of 99.92\% isotopically pure D$_2$O, contained in a 12-meter-diameter spherical acrylic vessel. This vessel was surrounded by 9456 20-cm Hamamatsu R1408 photomultiplier tubes (PMTs), which were installed on an 18 m diameter geodesic structure (PSUP). It is useful in this analysis to define the radius from the center of the detector to the front-face of the PMTs, $R_{\rm PMT}$ = 830 cm. A light concentrator \cite{Doucas1996579}  was mounted in front of each PMT to give a total photocathode coverage of nearly 55\% of $4\pi$. The acrylic vessel was surrounded by 7.4 kilotonnes of ultra pure H$_2$O. The acrylic vessel had a cylindrical section at the top, referred to as the neck, to allow deployment of calibration sources.

SNO was operated in three physics phases, which measured the total active solar neutrino flux with different techniques. The first operational phase (Phase I) used the deuteron as both the neutrino target and the neutron capture target for the neutrino-deuteron neutral-current (NC) measurement \cite{prlCC,prlNC}.  In Phase II, two tons of NaCl were added to the D$_2$O, which enhanced the efficiency of detecting the neutrons from NC interactions via radiative captures in $^{35}$Cl \cite{prlSALT}.  In Phase III, an array of proportional counters was deployed in D$_2$O \cite{SNO_NIMPaper,prc3Phase}. The proportional counters were constructed of approximately 2-m long high purity nickel tubes welded together to form longer strings. The array consisted of 36 strings filled with $^3$He, and an additional 4 strings filled with $^4$He that were insensitive to the neutron signals and were used for background studies.

In addition to solar neutrinos, SNO also studied atmospheric neutrinos \cite{PhysRevD.80.012001}. Since the $n$-$\bar{n}$ events are at the same energy scale as the atmospheric neutrino events, the data selection for this study of neutron-antineutron oscillations followed the same criteria as the SNO atmospheric neutrino analysis. The live times for the selected data are $350.43\pm0.01$ days in Phase I, $499.42\pm0.01$~days for Phase II and $392.56\pm0.01$ days for Phase III.   The total number of neutrons from deuterons contained in the spherical acrylic vessel was ($6.021\pm0.007$)$\times10^{31}$ in Phase I and II. The inclusion of the proportional counters reduced the overall number of neutrons in the D$_2$O to ($6.015\pm0.007$)$\times10^{31}$ in Phase III.

Since the neutron-antineutron oscillations signal is nucleus-dependent, the exposure of neutrons is categorized by nuclei:
\begin{eqnarray}
\mbox{neutron exposure (D)} &=& 2.047\times10^{32} \mbox{ n} \cdot \mbox{yr}\\
\mbox{neutron exposure ($^{16}$O)} &=& 8.190\times10^{32} \mbox{ n} \cdot \mbox{yr}
\end{eqnarray}
for the combined live time of all phases of SNO.  The analysis presented in this paper used a blind analysis: 50\% of the Phase I, 15\% of phase II  and 20\% of phase III data were made available to develop the reconstruction techniques and analysis criteria.


\section{Neutron antineutron Oscillations}\label{Section:NNBAR}

The suppression factor, $R$, is evaluated theoretically using the Paris potential for the case of the deuteron, and an optical potential for heavier nuclei \cite{PhysRevC.79.054001}.   In Dover, Gal and Richards \cite{Dover}, the average suppression factors in deuteron and in $^{16}$O were evaluated to be (2.48$\pm$0.08)$\times$10$^{22}$~s$^{-1}$ and ($10.0\pm2.0$)$\times$10$^{22}$~s$^{-1}$, respectively. Newer calculations from Friedman and Gal \cite{Gal}, based on more complete work on antiproton-nucleus interactions at low energies, evaluated a suppression factor for $^{16}$O of 5.3$\times$10$^{22}$~s$^{-1}$, about a factor of 2 lower than the previous estimate. A suppression factor for $^{56}$Fe was also evaluated to be a factor of 2 lower than Dover {\em et al}. These newer suppression factors improved both the Super-Kamiokande and the Soudan II experimental lower limits.  The suppression factor in the deuteron was not re-evaluated in their study due to inadequacy of the optical-potential approach for the deuteron \cite{Friedman2017}.

A new evaluation of the suppression factor in the deuteron was made by Kopeliovicha and Potashnikova that includes the spin dependence of the $\bar{n}p$ annihilation amplitudes and a reevaluation of the zero-range approximation of the deuteron wavefunction \cite{Kopeliovicha}:
\begin{equation}
 R_D \approx 2.94\times10^{22}\left(\frac{3r + 1}{4r}\right) \mbox{ s}^{-1}
\end{equation}
where $\sigma^{ann,S=1,0}_{\bar{n}p}$ are the triplet and singlet antineutron-proton annihilation cross section, respectively, and  $r=\sigma^{ann,S=1}_{\bar{n}p}/\sigma^{ann,S=0}_{\bar{n}p}$. In the limiting case where $r=1$, $R_D$ is 2.94~$\times$10$^{22}$~s$^{-1}$. In the case where $r\gg1$, the suppression factor is 2.21 $\times$10$^{22}$~s$^{-1}$. The allowable range of suppression factor is thus [2.21, 2.94] $\times$10$^{22}$ s$^{-1}$, consistent with the previous Dover {\em et al.} estimates.

Of the two specific cases relevant to this analysis, $n$-$\bar{n}$ oscillations in $^{16}$O and $^{2}$H, we have chosen to study only the latter due to SNO's low sensitivity to $^{16}$O. The reasons for this low sensitivity are explained in the next section.

\subsection{ \texorpdfstring{$n$-$\bar{n}$}{n-nbar} in \texorpdfstring{$^{16}$O}{16-O}}

An oscillated neutron ($\bar{n}$) in $^{16}$O may interact with the surrounding nucleons either through $\bar{n}n$ or $\bar{n}p$ interactions. The $^{16}$O Fermi momentum ($\sim$225 MeV) transferred to the daughter particles results in tracks that are closer in direction to each other than if the interaction had occurred at rest; this in turn complicates the reconstruction of the daughter particle's track.

The decay channels of an $n$-$\bar{n}$ oscillation in $^{16}$O are deduced from the final-state population of $\bar{n}p$ and $\bar{n}n$ collisions from beam experiments \cite{antineutronPhys}. The $\bar{n}n$ channels (not present in the deuteron) are more complex due to isospin\footnote{In the case of $\bar{n}p$ annihilation, only spin-1 interactions are involved, however for $\bar{n}n$ annihilation both spin-0 and spin-1 interactions are allowed.}.  Multiple daughters that mainly consist of charged and neutral pions populate these annihilation channels.

A further complication in the measurement of any of these channels comes from the interaction of the daughters with the immediate surrounding nucleons. According to Super-Kamiokande's studies \cite{nnbarSuperK}, the surrounding nuclear media absorb $\sim$23\% of the outgoing pions after an $\bar{n}p$  or $\bar{n}n$ interaction.

These two factors, the Fermi momentum transfer and the pion absorption, add significant uncertainties to the measurement of $n$-$\bar{n}$ oscillations in nuclear environments that are more complex than in the deuteron. The deuteron case is simpler and will be the focus of this paper. The inclusion of $^{16}$O in this analysis is estimated to give a less than 10\% improvement to the deuteron-only results.

\subsection{\texorpdfstring{$n$-$\bar{n}$}{n-nbar} in Deuteron}\label{SNO:Deuteron}

In the deuteron, only $\bar{n}p$ interactions are possible since no other surrounding nucleon exists. The lower average nucleon
Fermi momentum in the deuteron ($\sim$50 MeV) compared to that in $^{16}$O
results in daughter tracks that are more widely separated in direction.

\nnbarModels

The decay channels for $n$-$\bar{n}$ oscillations in the deuteron are deduced from the final-state population of neutron and antiproton collisions from beam experiments. There are measurements in two distinct momentum regimes that describe the daughter products from $\bar{n}p$ interactions:

\begin{itemize}
\item Momentum regime I (at rest) : A study of channels of an antiproton colliding with a neutron near rest showed a majority of 2-body intermediate states \cite{2body}.  These intermediate states can then decay into channels including multiple pions; however, the decay of the intermediate states is not constrained to pion-only final states.

\item Momentum regime II ($\sim$ 250 MeV) : Alternative interaction channels \cite{nnbarSuperK} for  $\bar{n}p$ annihilation have also been modeled using beam data of $\bar{p}n$ collisions at momenta comparable to the $^{16}$O Fermi momentum,  leading to an enlarged phase space for the proton-antineutron modes.
\end{itemize}

The $n$-$\bar{n}$ events in deuteron will fall in between these two regimes since the Fermi momentum ($\sim$50 MeV) is not at rest nor at 250~MeV. Figure \ref{NNBAREnergy} shows the visible light output of the different channels by which an $n$-$\bar{n}$ oscillation in the deuteron can be observed. Nearly all channels include multiple pions. Within the Momentum Regime I, heavier mesons, such as ($\rho,\omega,...$), will further decay and create more pions. Because of this, special attention is paid to the pion signature in this paper.

The visible light output is different between the two momentum regimes as can be observed in Figure \ref{NNBAREnergy}.  Both regimes are independently studied to understand the possible impact of this uncertainty on our analysis. As will be covered in Sec. \ref{sec:SystSect}, the average $n$-$\bar{n}$ detection efficiency is slightly different for the two momentum regimes and a weighted average of the efficiencies is used in the final analysis.


\section{Atmospheric Neutrino Backgrounds}\label{Section:AtmoNu}

Atmospheric neutrinos are the main background for searches such as proton decay and $n$-$\bar{n}$ oscillations. Energetic electrons, muons, or taus can be created by charged-current interactions, and if the neutrinos have enough energy, pions and other particles may also be created by resonance.  These pions and other particles form the background to the search of $n$-$\bar{n}$ oscillations.

The SNO detector response to these backgrounds is simulated in a three-step process. For the atmospheric neutrino flux the Bartol three-dimensional flux prediction \cite{PhysRevD.70.023006} is used and neutrino interactions are modeled by the NUANCE simulation package \cite{NUANCE}.  The output of NUANCE is then simulated in SNOMAN \cite{SNOFirst}, which evaluates the SNO detector response to these events.  Through-going events, defined as neutrino-induced muons created outside the detector volume that subsequently traverse the detector, are used to measure the atmospheric neutrino flux \cite{PhysRevD.80.012001}. The measured flux is $\phi_{norm}=1.22\pm0.09$ times higher than the Bartol prediction. The predicted overall flux of atmospheric neutrinos is scaled by this factor in this analysis. The through-going events, both simulated and measured, are only used as calibration of the event reconstruction algorithm.

A contained event is defined as an event that originated within the detector volume ($R<R_{\rm PMT}$) and whose progeny did not exit the detector.  The selected events for the analysis of $n$-$\bar{n}$ oscillation are required to be contained events.

The following types of contained events from atmospheric neutrino interactions are modeled by NUANCE:
\begin{eqnarray}
\label{eq:atmointeractions}
\nu_{\rm cc   }:&\nu_l N &\rightarrow l N\;\;\;\;\;\;\;\;\;\;\;\;\;\;\;\;\; \mbox{ Quasi-elastic CC}\nonumber\\
&\nu_l N &\rightarrow l N'\;\;\;\;\;\;\;\;\;\;\;\;\;\;\;\; \mbox{  Deep-inelastic CC}\nonumber\\
&\nu_l N &\rightarrow l N'\;\;\;\;\;\;\;\;\;\;\;\;\;\;\;\; \mbox{  Cabibbo-suppressed CC}\nonumber\\
\nu_{\rm nc   }:&\nu_l N &\rightarrow \nu_l N'\;\;\;\;\;\;\;\;\;\;\;\;\;\; \mbox{  Deep-inelastic NC}\nonumber\\
\nu_{\pi}:&\nu_l N &\rightarrow l \Delta \rightarrow l N' \pi \;\;\;\;\mbox{  CC pion creation}\nonumber\\
&\nu_l N &\rightarrow \nu_l \Delta \rightarrow \nu_l N' \pi\; \mbox{  NC pion creation}\nonumber\\
\nu_{X    }:&\nu_l N &\rightarrow l(\nu_l)  X \;\;\;\;\;\;\;\;\;\;\;\;\mbox{  CC(NC) $n\pi$}\nonumber\\
\nu_{\rm otr }:&\nu_l N &\rightarrow l(\nu_l)  X \;\;\;\;\;\;\;\;\;\;\;\;\mbox{  ES, IMD, PNP}
\end{eqnarray}
where $l=\{e,\mu,\tau\}$, $N=\{p,n\}$ and $X=\{\rho,\eta,\Sigma,...\}$ (which in many cases decay into pions), ES refers to elastic scattering , IMD to inverse muon decay, and PNP to photonuclear production ($\nu_{l}N\rightarrow N l \gamma$). Charged pions originating from atmospheric neutrino interactions in the detector, either through $\Delta\pi$ resonance or exotic particle creation, are an irreducible source of backgrounds to the $n$-$\bar{n}$ oscillations search in this analysis.

\xnoedFigures

\section{Event reconstruction algorithm}\label{Section:PIDReco}

To reconstruct the $n$-$\bar{n}$ signal, it is necessary to understand both the individual signature of each pion daughter and the overall signature of simultaneous particles propagating in the detector. In water Cherenkov imaging detectors there are two distinctive signatures, ``showering'' and ``non-showering'', that indicate whether a particle cascade has occurred or not.  These signatures are used for particle identification.    A naming convention of $e$-like (showering) and $\mu$-like (non-showering) is used in this context since these signatures distinguish between electrons and muons in atmospheric neutrino interactions, which form the majority of the backgrounds.  The $n$-$\bar{n}$ daughters are composed of pions. The charged pions have a different signature than the neutral pions; the neutral pions have an $e$-like signature, while the charged pions have a signature that is more complex, as will be described in Sec.~\ref{PionSection}.

A relativistic charged particle emits Cherenkov photons along its track at an angle relative to the track direction of $\theta_c\leq 41.2^\circ$ in the D$_2$O.  The topology of the PMTs that have been triggered by these photons resemble a circular ring. Single-ring events are illustrated in Fig.~\ref{CherenkovSignature} for the two signatures that are distinguishable in SNO. Multiple-ring events appear from interactions (e.g. the hard scattering of a particle) or particle decays (e.g. pion decays) with multiple progenies in the final state. The correct identification of multiple-ring events is necessary in the search of $n$-$\bar{n}$ oscillations.

The spatial location of each PMT is expressed in spherical coordinates ($\cos\theta_{\rm PMT}$, $\phi_{\rm PMT}$, $R_{\rm PMT}$) with the center of the detector as the origin. Since the radial component of the spherical coordinate is fixed at $R_{\rm PMT}$, the hit PMT pattern can be displayed in the ($\cos\theta_{\rm PMT}$, $\phi_{\rm PMT}$) space (see Fig.~\ref{CherenkovSignature}) as a two-dimensional image, allowing the use of two-dimensional pattern-finding techniques. SNO developed its own pattern-finding algorithm for these images.

In this analysis, ring counting is done via a Multiple Ring Fitter (MRF) \cite{PHD}. This fitter is composed of four parts: (1)~an algorithm to search for possible rings/circles in an image created by the hit PMT pattern  of the event, (2)~an algorithm to sort the possible rings/circles into 12 distinct regions, (3)~an algorithm to estimate the position and direction of possible particles in the 12 distinct regions, and (4)~an algorithm to validate the rings and to deduce the corresponding signature.  These four parts are discussed below.

(1)~A single ring can be parametrized with ($\cos\theta_o$, $\phi_o$, $\rho$); two ring-centered angular coordinates ($\cos\theta_o$, $\phi_o$) and a ring arc radius $\rho$ defined as the radius of the ring constrained to the spherical surface of the detector.  This ring-parameter space ($\cos\theta_o$, $\phi_o$, $\rho$) is used as a likelihood space --- or Hough space~\cite{HoughRef} --- for the detection of rings.  Each pair of triggered PMTs is mapped into this ring-parameter space. The first two parameters that describe the triggered pair in the ring space are the coordinates ($\cos\theta_{mp}$, $\phi_{mp}$) of the midpoint between the two PMTs.  The other parameter is the arc length from the midpoint to one of the PMTs in the pair.  Each point in this ring-parameter space defines a possible ring of a certain radius at a certain point on the surface of the detector.

(2)~After all pairs of triggered PMTs have been mapped to this space, the point in the ring-parameter space with the highest density is considered the most likely ring candidate.  In the case of multiple rings, there will be a series of local high-density maxima across the ring-parameter space.  A further sub-division of this ring-parameter space is made to look for these local maxima. This is done using the sub-sections of a dodecahedron, constructed to approximate a sphere with 12 pentagonal surfaces (see Fig.~\ref{Dodecahedron}). Each of these surfaces is considered an independent likelihood space leading to a possibility of a total of 12 rings in an  event.  The dodecahedron structure is rotated so that the center of the best possible ring is at the center of one of the pentagonal surface; this ring is considered the primary ring. This rotation allows better ring separation.
\dodecahedron

(3)~The vertex and track of the particle that produced the ring is reconstructed by assuming a Cherenkov light cone with an opening angle of 41.2$^\circ$.  The track reconstruction is complicated by the spherical nature of the SNO detector. It can be  shown that for the spherical geometry of SNO, the ring pattern is mostly circular, independent of the vertex location, and as such the most complete likelihood function for an accurate reconstruction would require PMT timing information.  In this analysis, the timing information is not included due to the complexity of the time structure of high-energy events caused by light reflection at the acrylic vessel. A reconstruction algorithm without the incorporation of the PMT timing structure is found to be adequate for this analysis.  It is assumed that the direction of the reconstructed track $\hat{u}_{rec}$ for each ring follows
\begin{equation}
\hat{u}_{rec}= \sin\theta^{\dagger}_{mp}\cos\phi^{\dagger}_{mp} \hat{x}+ \sin\theta^{\dagger}_{mp}\sin\phi^{\dagger}_{mp} \hat{y}+ \cos\theta^{\dagger}_{mp} \hat{z},
\end{equation}
where $\theta^{\dagger}_{mp}$ and $\phi^{\dagger}_{mp}$ denote the point in the sub-divided ring-parameter space with the highest density.  Each ring is then considered to have its own reconstructed vertex $x'_{rec}\hat{u}_{rec}$.

While the omission of the timing information increased the uncertainty in vertex reconstruction accuracy, this analysis is concentrated on counting the total number of rings in an event and does not rely on the precise knowledge of the reconstructed vertex (see Sec.~\ref{PionSection}).
\AngularDistribution

(4)~ An additional verification method is implemented for each ring candidate, $\alpha$.  Each fired PMT, $i$, is transformed into an opening angle $\xi_i$; $\xi_i$ is defined as the angle subtended by the vector from the fitted vertex to the ring center coordinate ($\cos\theta_{mp}^{\dagger}$, $\phi_{mp}^{\dagger}$) and the vector from the fitted vertex to the position of the fired PMT $i$. Each $\xi_i$ is collected in a binned histogram containing 30 bins in the range of 0$^\circ$ to 60$^\circ$. Once each $\xi_i$ is collected, the resulting distribution is compared to the expected distribution for either the showering or the non-showering signature $\xi_{exp}$ (shown on Fig.~\ref{Angulardistribution}) using a likelihood method. The behavior of the two distributions at opening angles larger than the Cherenkov opening angle (41.2$^\circ$) differs and this difference allows good separation between the two signatures.

The likelihood for each candidate is evaluated over all bins with
\begin{equation}\label{logLambda}
-2 \ln \lambda_{\alpha}= \\  2\sum_j \left[(\xi^{exp}_j-\xi_j)+\xi_j\ln (\xi_j/\xi^{exp}_j )\right],
\end{equation}
where $j$ is the index of the bin.

It was observed in simulations that the electron-ring expectation was able to identify rings for both electrons and muons with high confidence. The muon-ring expectation only identified rings originating from muons, but proved less efficient at identifying these rings compared with the electron-ring expectation.

\subsection{Reconstruction of \texorpdfstring{$\pi^\pm$}{pi-plus-minus}}\label{PionSection}

Our ability to correctly identify a ring of simulated charged pions with an electron-ring expectation proved to be more efficient by an order of magnitude compared to when we used a muon-ring expectation. For charged pions of 600 MeV, we can correctly identify the primary ring 68\% of the time with an electron-ring expectation, while we could only identify the primary ring 12\% of the time with a muon-ring expectation. This is directly tied to the signature of pions in a heavy water Cherenkov detector, which is different from the signature of either a muon or an electron. Charged pions should have a non-showering signature, but two competing processes complicate the reconstruction of the track and local vertex:
\begin{itemize}
\item The charged pions' short lifetime of 26 ns can interrupt the production of light along the track. The outgoing $\mu^\pm$s will generally follow a different track after the $\pi^{\pm}$ decays, thus increasing the probability of observing multiple rings within the detector, or they can be below the Cherenkov threshold.
\item The direction of the outgoing particle changes after an elastic or inelastic scattering producing additional tracks; in some cases additional pions may be produced and propagate in the detector.  Inelastic processes include single charge exchange (SCX), double charge exchange (DCX), $\Delta\pi$ resonance and absorption.
\end{itemize}
The  mean free path for $\pi^{\pm}$ to undergo an inelastic interaction is shorter than the typical range over which the pion loses all of its energy. In D$_2$O, the mean free path for nuclear interactions is about three times lower than in H$_2$O.  These pion inelastic interactions are studied using the Bertini cascade model, incorporated into a corrected Hadron-CALOR model that has been integrated within SNOMAN \cite{Calor,PHD}. It is important to note that while in the case of H$_2$O the dominant process is $\pi^-$ absorption on $^1$H, there is no difference in cross section between $\pi^-$ and $\pi^+$ for absorption in the deuteron.

\piplusReconstructionFigures

Figure \ref{fig:piplus_mu_reco} show scatter plots of the reconstructed position of pions generated at the center of the detector and the angle between the particle's original direction and the final reconstructed direction using showering and non-showering expectation. Events at the coordinate ($x'_{rec}$=0, $\theta_{rec}$=0) are events that correctly reconstruct both the original vertex and direction of the particle.

If the charged pion undergoes a SCX, the resulting $\pi^0$ will decay and will require an $e$-like ring expectation to fit the two outgoing rings.  A visual inspection of the reconstructed ring shows that the technique is sound for rings with $x'_{rec}< 350$ cm  along $\hat{u}_{rec}$.

The reconstruction fails for rings with reconstructed position of $x'_{rec}> 350$ cm  along $\hat{u}_{rec}$ due to low statistics when the ring is close to the edge of the detector as is shown in Fig.~\ref{small-ringPathology}.

\xnoedPatho

The cost of using a showering-ring expectation to detect non-showering rings is inferred from Fig.~\ref{fig:piplus_mu_reco}.  A muon reconstructed with a showering expectation suffers a bias in the reconstructed vertex position and the track direction, leading to a smaller detected ring.  In the context of this analysis, i.e.~counting the number of rings in an event, the accurate reconstruction of event vertex position is not required, and ring counting has been proved adequate.

The precise reconstruction of a single charged pion is not the goal of this analysis, since our signal is composed of many pions being generated at once. Multiple-ring  Monte Carlo simulation studies showed that the difference in topology between the signal and the background is distinctive enough to separate $n$-$\bar{n}$ multiple-pion events from a wide class of atmospheric neutrino interactions.

It is important to emphasize that the ring pattern created by the $n$-$\bar{n}$ oscillation signal, formed of charged and neutral pions, is reconstructed solely using electron-ring expectations in this analysis.  The performance of the MRF is benchmarked on GeV-scale energy through-going (i.e.~not contained) events and comparisons of simulated and actual data sets are used to evaluate systematic uncertainties on the fitter.


\section{Analysis}\label{event:Selection}
\subsection{Analysis Parameters}

\begin{table*}
\centering
{\footnotesize
\caption[Summary of analysis parameters, data selection and noise rejection criteria.]{Summary of analysis parameters, data selection and noise rejection criteria.}
\label{ParaTable}
\begin{tabular}{   l  l}
\hline\hline

\multicolumn{2}{l}{{{\it \bf Analysis Parameter 1: Multiple Cherenkov rings acceptance criterion}}}\\
$N_{\rm rings}>1$ & Two or more Cherenkov rings reconstruct in a single event \\

\ \ \ $\cos\theta_{\rm ring}<0.86$ & Minimal angle at which two rings are distinguishable \\

\ \ \ $x'_{rec}<350$ cm & Minimal size of Cherenkov rings. Implemented to remove false positives as described\\
& in Sec.~\ref{PionSection}\\

\multicolumn{2}{l}{{{\it \bf Analysis Parameter 2: Isotropy acceptance criterion}}}\\
$\Lambda<m_m\cdot p.e.+b_m$ & The vector sum of the total light in each ring (Eqn. \ref{Lambda}) is below an energy dependent \\
&threshold. See text for details on $\Lambda$. \\

\multicolumn{2}{l}{{{\it \bf Additional event selection criteria}}}\\
Contained cut &  Select events that originate and terminate in the detector by that less than 4 veto \\
& PMTs registered light out of 91 potential veto PMTs\\

$2000<p.e.<18000$ & Visible energy (the number of photoelectrons, $p.e.$) of the event  approximately \\
&corresponding to an energy window of 250 MeV to 2 GeV. \\
\hline\hline

\multicolumn{2}{l}{{{\it \bf Instrumental background event rejection}}}\\

Neck events & Remove instrumental backgrounds that originate from the ``neck" region, place where \\
& the acrylic vessel is connected\\

Retrigger & Remove events that occurred within a 5 $\mu s$-window
 prior to the current event.\\

Burst & Remove instrumental high-energy events that trigger the detector in quick succession;\\
&if four or more successive non-retrigger events are tagged within a 2-second period,\\
&all the events are removed.\\

Pmt$\_$hit/Nhits$>0.7$ & Ratio of PMTs with good calibration (Pmt$\_$hit) over all triggered PMTs (Nhits)\\
 & for the event.\\
\hline\hline
\end{tabular}
}
\end{table*}

A summary of the analysis parameters and data selection criteria are presented in Table~\ref{ParaTable}. This analysis relies on two parameters evaluated on contained events.  The first parameter is the number of detected Cherenkov rings, $N_{\rm rings}$, and the other is an isotropy estimator for multiple-ring events.  These two parameters are efficient in isolating $n$-$\bar{n}$'s multiple-pion signal from atmospheric neutrino backgrounds.

Figure~\ref{MultRing} shows the expected $N_{\rm rings}$ distributions for the atmospheric background and the $n$-$\bar{n}$ signal compared to the data. Two additional ring selection conditions are applied. The first condition eliminates rings whose tracks are too close to one another.  In multiple-ring event candidates, the ring with the lowest $-2\ln\lambda$~(Eqn.~\ref{logLambda}) is kept.  At $\cos\theta_{\rm ring}<0.86$ two distinctive rings could be separated, where $\cos\theta_{\rm ring}$ is the angle between the tracks of two reconstructed rings.   The second condition eliminates rings that are too small due to false reconstruction from the fitter, as described in Sec.~\ref{PionSection}, by imposing the criterion of $x'_{\rm rec}<350$~cm. There is good agreement between data and the atmospheric neutrino expectation for this analysis parameter.

The analysis is further refined by taking advantage of the isotropic nature of $n$-$\bar{n}$ interactions. An energy-dependent isotropy selection is applied to multiple-ring events such that $\Lambda<m_m\cdot p.e.+b_m$, where $p.e.$ is the number of detected photoelectrons, and $m_m$ and $b_m$ are parameters derived from simulations.  The selection parameter $\Lambda\equiv|\vec{\Lambda}|$, which is a proxy for the total-momentum-to-energy ratio of an event, is defined as:
\begin{equation}
\vec{\Lambda}=\sum_{i=1}^{N}{\frac{\mbox{PMT}^{\;i}_{\rm ring}}{\mbox{Nhits}}} \cdot\hat{x}^{\;i}_c,
\label{Lambda}
\end{equation}
where PMT$^{\;i}_{\rm ring}$ is the total number of hit PMTs in a cone opening angle of 60$^{\circ}$ around the reconstructed track, and $\hat{x}^{\;i}$ is the direction of track $i$.

\RingAndPoverEFigures

The choice of the opening angle selection is based primarily on the fact that most prompt Cherenkov photons are detected below an angle of 41.1$^\circ$ for non-showering particles and below an angle of 60$^\circ$ for showering particles. Photons that are detected above this angle are either Rayleigh scattered, reflected by the acrylic vessel or belonged to another ring.

The $m_m$ and $b_m$ parameters are derived by studying the properties of atmospheric neutrinos and $n$-$\bar{n}$ oscillation simulations across the three operational phases in the $p.e$-$\Lambda$ parameter space. This selection boundary, seen in Fig.~\ref{3PhaseIso} for this combined 3-phase analysis, improved the signal-to-background separation. The optimized parameters are $m_m=-4.59\times10^{-5}$ and $b_m=0.921$, respectively.

Also presented in Table~\ref{ParaTable} are the criteria to isolate contained events from the cosmic through-going muon events and instrumental background events.
Good agreement is observed between the simulation and the selected data rates.  An analysis was performed and showed that the instrumental backgrounds expected were negligible: 0.18$\pm$0.13 instrumental events are expected after all analysis cuts have been applied \cite{PHD}.

\EffAndSysTable


\subsection{Efficiencies and Systematic Uncertainties}
\label{sec:SystSect}

The $n$-$\bar{n}$ signal acceptance and the contained atmospheric-neutrino contamination obtained after applying the two high-level cuts on the three phases of SNO data are shown in Tables~\ref{nnbarChan} and~\ref{atmoChan}, respectively. The last column shows the efficiency for a specific channel $i$ such that
\begin{equation}
\epsilon^{i}_{\rm tot}= \epsilon^{i}_{\rm multi-ring}\cdot\epsilon^{i}_{\rm isotropy\: cut}.
\end{equation}
The total signal detection efficiency (shown in bold in the table) is weighted by the channels' branching ratios $\Gamma^{i}$:
\begin{equation}
\epsilon_{\rm regime}=\frac{\sum_{i}\Gamma^{i}\epsilon^{i}_{\rm multi-ring}\cdot\epsilon^{i}_{\rm isotropy\: cut}}{\sum_{i}\Gamma^{i}}.
\end{equation}
In Table~\ref{nnbarChan}, the error on the signal detection efficiencies includes the differences of the detector response for the three  operational phases of SNO. The total $n$-$\bar{n}$ detection efficiency of Regime~I and~II are consistent.

The efficiency of observing an $n$-$\bar{n}$ event is ($54.0 \pm 4.6$)\% when averaging over all operational phases of SNO. This efficiency combines two physical regimes described in Section \ref{SNO:Deuteron} in a weighted average. A relative increase of 1.4\% in signal detection efficiency is observed in Phase~II compared to Phase~I.  This can be explained by the increase of the nuclear cross section in $^{35}$Cl for $\pi^{\pm}$. A relative decrease of 7.5\% in the detection efficiency of the signal is observed between Phase~I and Phase~III; this is caused by the optical shadowing of the NCD array, which impacted the light isotropy of the event.

The systematic uncertainties of these efficiencies and those of the selection criteria from the previous section are summarized in Table~\ref{table:SystErr}.  Also included in this table are the uncertainties of the neutrino oscillation parameters, the uncertainties associated with the atmospheric neutrino flux, and uncertainties in atmospheric-neutrino interactions at the GeV-scale.

The total systematic uncertainty on the detection of the $n$-$\bar{n}$ oscillation signal ($\sigma_{signal}$) in the full SNO data set is
\begin{equation}
\sigma_{signal}=
\frac{\sum_{i}T_{i}\sigma_{\epsilon_{i}^{n-\bar{n}}}}{\sum_{i} T_{i}}
\end{equation}
where $\sigma_{\epsilon_{i}^{n-\bar{n}}}$ is the uncertainty associated with the detection efficiency and $T_{i}$ is the live time of phase $i$. The total systematic uncertainty on the $n$-$\bar{n}$ detection efficiency is found to be 11.7\%. This is dominated by the $\bar{n}p$ modeling, which combines the systematic uncertainty on the $\bar{n}p$ branching ratio for models in Sec.~\ref{SNO:Deuteron} and the efficiency of observing a $n-\bar{n}$ from Table~\ref{nnbarChan}.

The total systematic uncertainty on the expected background from atmospheric-neutrino interactions ($\sigma_{\rm bkgd}$) is a combination of the uncertainties on neutrino oscillation parameters ($\sigma_{\Phi_{\rm osc}^{\rm atmo}}$), the uncertainties on detector efficiencies due to cut parameters and MRF reconstruction of through-going muons
($\sigma_{\epsilon_{i}^{\rm atmo}}$), and the uncertainties in the amplitude of different atmospheric-neutrino interaction channels ($\sigma_{\Phi_{i}^{\rm atmo}}$):
\begin{equation}
\sigma_{\rm bkgd}=
\sqrt{\sigma_{\Phi_{\rm osc}^{\rm atmo}}^2
\!\!+\!\!
\left(\frac{\sum_{i}T_{i}\sigma_{\Phi_{i}^{\rm atmo}}}{\sum_{i} T_{i}}\right)^2
\!\!+\!\!
\left(\frac{\sum_{i}T_{i}\sigma_{\epsilon_{i}^{\rm atmo}}}{\sum_{i} T_{i}}\right)^2}
\end{equation}
and is 24.5\% for the 3-phase SNO data set.

No calibration sources were available to calibrate events in the GeV scale. Through-going muon events are used to characterize the fitter response between data and Monte Carlo.  Since these events originated from outside the detector, there remain uncertainties in the response of the detector to contained events. A shift in likelihood space is observed for the multiple-ring fitter response between Monte Carlo and data of these through-going ``calibration'' events. Since we apply a cut on the likelihood space to determine whether there is a ring or not, there is a systematic uncertainty associated with this shift. This is referred to as the MRF reconstruction calibration uncertainty.

In the case of atmospheric neutrino detection efficiency of contained events, the studies of the MRF calibration led to a systematic uncertainty of $\sim$15\% in the observed number of rings.  The effect of the MRF calibration is not noticeable as much for the $n$-$\bar{n}$ oscillation detection efficiency with an uncertainty due to the calibration of $\sim$7\%, in part due to cleaner reconstruction of the individual rings. The specific uncertainties due to MRF calibration for each phase of SNO are included in Table~\ref{table:SystErr}.

Compared to the aforementioned detector response and reconstruction uncertainties, those associated with neutrino oscillation parameters are negligible in this analysis.  Other sub-dominant uncertainties in the total atmospheric background include the Bartol atmospheric flux normalization (7.4\%) and the production rate of pions from $\Delta$ resonance (8\%).

\resultTable
\section{Results}\label{Section:Results}

Table~\ref{3PhasesResults} shows the results after applying all data selection criteria to the data of the three SNO operational phases; also shown are the expected backgrounds for the three phases.

For the combined 3-phases analysis, the observed result of 23~events is 1.6 $\sigma$ lower than the expected background of 30.5~events, which is consistent with statistical fluctuation.  The number of observed events and expected background are also statistically consistent in each phase.

This result is transformed into a lower limit on $n$-$\bar{n}$ oscillation lifetime as is discussed in the following section.

\subsection{Limit Evaluation and Neutron-antineutron Lifetime}\label{nnbarLIFE}
\LimitResultTable

The profile likelihood method \cite{Rolke} introduces a way to include systematic uncertainties in the evaluation of confidence intervals in rare-event searches.  In this analysis, the technique is employed with Gaussian errors, both for the background rate and for detection efficiency. Other systematic effects, such as instrumental backgrounds, are negligible. The likelihood function is given by
\begin{equation}
L(\mu,b,\epsilon | x,b_{o},\epsilon_{o})=P_{P}(x|\mu,\epsilon,b) P_{G}(\epsilon|\epsilon_0,\sigma_{\epsilon})P_{G}(b|b_0,\sigma_b)
\end{equation}
where $P_P$ and $P_G$ are respectively the Poisson and Gaussian probability density function; $\mu$ is the signal rate of the rare event, $b$ is the background rate and $\epsilon$ is the signal detection efficiency; $x$ is the observed number of events, $b_o$ and $\epsilon_o$ are the expected value for the background rate and efficiency.
The likelihood ratio $\mathcal{L}$, used to evaluate the confidence interval, consists of the supremum of the likelihood function
\begin{equation}
\mathcal{L}(\mu | x,b_{o},\epsilon_{o}) = \frac{\mbox{sup}(L(\mu,\hat{b}(\mu),\hat{\epsilon}(\mu) | x,b_{o},\epsilon_{o}))}{\mbox{sup}(L(\hat{\mu},\hat{b},\hat{\epsilon} | x,b_{o},\epsilon_{o}))}
\label{profLikeEqu}
\end{equation}
where $\hat{b}$, $\hat{\epsilon}$ or $\hat{\mu}$ are values that maximize the likelihood function. When a measured signal is less than the expected background, an issue arises with the ${\mbox{sup}(L(\hat{\mu},\hat{b},\hat{\epsilon} | x,b_{o},\epsilon_{o}))}$ part of Eqn. \ref{profLikeEqu}, as $\hat{\mu}$ can become negative.  If it is allowed to be negative the limit is said to be unbounded (UB); while if it is not, the denominator is evaluated at sup($L(\hat{\mu},\hat{b},\hat{\epsilon} | x,b_{o},\epsilon_{o}))|_{\hat{\mu}=0}$ and the limit is said to be bounded (B).

In the process of evaluating the confidence interval in this analysis, studies were performed to verify the statistical coverage of the technique \cite{PHD}.  The unbounded technique offers better coverage and is set as default in the TRolke~2.0 package \cite{Rolke2}, which is used in this analysis. However, it is customary to use a bounded limit and both results will be presented in this article.

The free oscillation time $\tau_{\rm free}$ is evaluated as
\begin{equation}
\tau_{\rm free}=  \sqrt{T_{\rm intranuclear}\cdot \left(\frac{3.16\times 10^{7}  \mbox{ s/year}}{R}\right) } \label{exposure}
\end{equation}
where
\begin{equation}
T_{\rm intranuclear} = \frac{\mbox{exposure}\times\epsilon_{n-\bar{n}}}{\mbox{UL}}
\end{equation}
and UL is the upper limit of the signal evaluated at 90\% CL.  For the deuteron-only scenario, the nuclear $T_{\rm intranuclear}$ limit is evaluated at {1.18 $\times10^{31}$(bounded) or 1.48$\times10^{31}$ (unbounded)} years at 90\% CL. Shown in Table \ref{nnbarLimit2} are the limits evaluated with the profile likelihood method for different operational phases of SNO.

\subsection{Discussion of results}\label{sec:Discussion}

As is shown in Fig.~\ref{MultRing}, the number of multiple-ring events prior to the isotropy cut agrees well with the expected number of multiple-ring events from the atmospheric neutrino background.  The isotropy cut is efficient in separating $n$-$\bar{n}$ from a wide class of atmospheric neutrino interactions as shown on Fig.~\ref{3PhaseIso}.

For the deuteron-only scenario, the nuclear lower limit presented above translates into a $\tau_{\rm free}$ limit of 1.2$\times10^{8}$~s (bounded) or 1.4$\times10^{8}$~s (unbounded).  We have chosen not to include $^{16}$O targets for reasons discussed in Section \ref{Section:NNBAR}, mainly the complexity associated with intranuclear effects in oxygen, but we estimate the improvement in the limit to be minimal.

 In order to compare our result to that from Super-Kamiokande, which uses a different method to evaluate a limit, we have re-evaluated their limit with the technique presented in this paper. The Super-Kamiokande intranuclear lifetime is re-evaluated with the profile likelihood method at {22.0} $\times10^{31}$~years (bounded and unbounded) at 90\% confidence level instead of {19} $\times10^{31}$~years. Since Super-Kamiokande observed 24 events and expected 24.1 $\pm$ 5.7, the difference between a bounded and unbounded limit is minimal.  This represents a 16\% difference compared to the published SK limit, which translates to a free oscillation lower limit of $2.9\times10^{8}$~s at 90\% CL.

Future improvements on the $n$-$\bar{n}$ oscillations search could be obtained by requiring the measurement of multiple Michel electrons emerging from the $\mu^\pm$ daughters of $\pi^\pm$ decays.  While the trigger window in most water Cherenkov detectors is too large to study $\pi$ decays (26 ns lifetime), it is possible to study the lifetime of $\mu$ decays in separate events. This indirect tagging of $\pi$ may be beneficial for removing atmospheric neutrino backgrounds since multiple Michel electrons may only be present in $\sim11\%$ of total contained atmospheric neutrinos.


\section{Conclusion}\label{sec:Conclusion}
In this paper a new nuclear limit on the neutron-antineutron oscillation search in deuteron is obtained.  The intranuclear oscillation life time is 1.18$\times 10^{31}$~(bounded) or 1.48 $\times 10^{31}$~(unbounded) years at 90\% CL from the data of all three SNO operational phases. This translates into a free oscillation limit of 1.23$\times10^{8}$ s ~(bounded) or 1.37$\times10^{8}$ s~(unbounded)~at 90\% CL using the models from Dover {\it et al}.  This is the first search for neutron-antineutron oscillations using deuterons as target.

\begin{acknowledgments}
This research was supported by: Canada: Natural Sciences and Engineering Research Council, Industry Canada, National Research Council, Northern Ontario Heritage Fund, Atomic Energy of Canada, Ltd., Ontario Power Generation, High Performance Computing Virtual Laboratory, Canada Foundation for Innovation, Canada Research Chairs program; US: Department of Energy, National Energy Research Scientific Computing Center, Alfred P. Sloan Foundation; UK: Science and Technology Facilities Council (formerly Particle Physics and Astronomy Research Council); Portugal: Funda\c{c}\~{a}o para a Ci\^{e}ncia e a Tecnologia. We thank the SNO technical staff for their strong contributions.  We thank INCO (now Vale, Ltd.) for hosting this project in their Creighton mine.
\end{acknowledgments}


\end{document}